\documentclass[twocolumn,epjc3]{svjour3}  

\smartqed  
\RequirePackage{graphicx}
\RequirePackage[numbers,sort&compress]{natbib}
\RequirePackage[colorlinks,citecolor=blue,urlcolor=blue,linkcolor=blue]{hyperref}
\journalname{Eur. Phys. J. C}

\usepackage{times}
\usepackage{float}
\usepackage{subfigure}
\usepackage{array}
\usepackage{amsmath}
\usepackage{amssymb}
\usepackage{amsfonts}
\usepackage{bm}
\usepackage{multirow}
\usepackage{booktabs}
\usepackage{dcolumn}
\usepackage{latexsym}
\usepackage{verbatim}
\usepackage{physics}
\usepackage[english]{babel}
\usepackage{widetext}
\usepackage[utf8]{inputenc}
\usepackage{todonotes}
\usepackage{epstopdf} 
\usepackage{microtype}

\usepackage{soul}

\graphicspath{{Figures/}}

\begin{document}

\title{Euclidean Thermodynamics and Lyapunov Exponents of Einstein-Power-Yang-Mills AdS Black Holes}

\author{ Karthik R \thanksref{addr1,e1}, 
Dillirajan D. \thanksref{addr1,e2}, 
K. M. Ajith \thanksref{addr1,e3}, 
Kartheek Hegde \thanksref{addr1,e4}, 
Shreyas Punacha \thanksref{addr2,e5}, 
A. Naveena Kumara \thanksref{addr3,e6}}

\thankstext{e1}{e-mail: raokarthik1198@gmail.com}
\thankstext{e2}{e-mail: dillirajan708090@gmail.com}
\thankstext{e3}{e-mail: ajith@nitk.ac.in}
\thankstext{e4}{e-mail: hegde.kartheek@gmail.com}
\thankstext{e5}{e-mail: shreyasp444@gmail.com, shreyas4@uw.edu}
\thankstext{e6}{e-mail: naviphysics@gmail.com, nathith@irb.hr}

\institute{Department of Physics, National Institute of Technology Karnataka, Surathkal 575 025, India \label{addr1} 
\and Department of Oral Health Sciences, School of Dentistry, University of Washington, Seattle, WA 98195, USA \label{addr2}
\and Theoretical Physics Division, Rudjer Bo\v skovi\'c Institute, Bijeni\v cka c.54,  HR-10002 Zagreb, Croatia \label{addr3} 
}

\date{Received: date / Accepted: date}

\maketitle

\begin{abstract}
We study the thermodynamics of Einstein-Power-Yang-Mills AdS black holes via the Euclidean path integral method, incorporating appropriate boundary and counterterms. By analyzing unstable timelike and null circular geodesics, we demonstrate that their Lyapunov exponents reflect the thermodynamic phase structure obtained from the Euclidean action. Specifically, the small-large black hole phase transition, analogous to a van der Waals fluid, is signaled by a discontinuity in the Lyapunov exponent. Treating this discontinuity as an order parameter, we observe a universal critical exponent of $1/2$, consistent with mean-field theory. These results extend previous insights from black hole spacetimes with Abelian charges to scenarios involving nonlinear, non-Abelian gauge fields, highlighting the interplay between black hole thermodynamics and chaotic dynamics.
\end{abstract}

\keywords{Einstein-Power-Yang-Mills Black Holes;  Euclidean Thermodynamics; Phase Transitions; Lyapunov Exponents.}

\maketitle

\section{Introduction}

Black hole thermodynamics provides a profound connection between gravity, quantum theory, and statistical mechanics \cite{Witten:2024upt}. The pioneering works of Bekenstein and Hawking in the 1970s established that black holes possess entropy and emit thermal radiation, forming the basis for the laws of black hole thermodynamics \cite{Hawking:1971tu, Bekenstein:1972tm, Bekenstein:1973ur, Bardeen:1973gs, Hawking:1974rv, Hawking:1975vcx}. According to this paradigm, black holes behave as thermodynamic objects characterized by well-defined temperature and entropy. Black holes in anti-de Sitter (AdS) spacetime exhibit rich phase structures analogous to ordinary thermodynamic systems \cite{Hawking:1982dh, Witten:1998zw}. Charged AdS black holes, such as the Reissner-Nordström AdS solution, display even closer analogies with familiar fluid systems \cite{Chamblin:1999hg, Chamblin:1999tk, Caldarelli:1999xj}. Specifically, for fixed charge, a first-order phase transition occurs between small and large black hole phases, reminiscent of the liquid-gas transition observed in van der Waals fluids. This analogy was subsequently made quantitative through the framework known as black hole chemistry, an extended thermodynamic approach where the negative cosmological constant \( \Lambda \) is interpreted as thermodynamic pressure \( P \), and its conjugate quantity serves as the black hole’s thermodynamic volume \cite{Kastor:2009wy, Dolan:2011xt, Kubiznak:2012wp, Gunasekaran:2012dq, Kubiznak:2016qmn}. This expanded \emph{chemical} perspective has introduced new terminology, such as small and large black hole phases, along with chemical analogies like triple points and reentrant phase transitions, regularly discussed in contemporary literature \cite{Altamirano:2013ane, Altamirano:2013uqa, Frassino:2014pha, Kubiznak:2016qmn, NaveenaKumara:2019nnt, NaveenaKumara:2020biu}.

In recent years, significant connections have emerged among black hole physics, chaos theory, and information theory. A crucial diagnostic of chaos is the Lyapunov exponent, which measures the rate at which nearby trajectories in phase space diverge or converge \cite{doi:10.1080/00207179208934253}. A positive Lyapunov exponent indicates sensitive dependence on initial conditions (chaotic behavior), while a negative exponent implies stable, regular motion. Remarkably, it has been observed that motion near a black hole horizon frequently saturates a proposed universal upper bound for the Lyapunov exponent, \( \lambda_{\max} = 2\pi T_{\mathrm{BH}} \) (in units where \( \hbar = 1 \)). This bound, initially conjectured in the context of gauge/gravity duality, suggests that black holes represent \emph{maximally chaotic} systems \cite{Maldacena:2015waa}. However, certain exotic scenarios have provided counterexamples, demonstrating violations of this bound and indicating the subtle nature of the relationship between gravity and chaos \cite{Zhao:2018wkl, Guo:2020pgq, Gwak:2022xje}. Interestingly, there is also a noted correspondence between the Lyapunov exponent governing unstable null geodesics (such as those defining the photon sphere of a black hole) and the imaginary component of certain quasinormal mode frequencies \cite{Cardoso:2008bp, Guo:2021enm}. Quasinormal modes characterize how black holes return to equilibrium following perturbations, with their frequencies encoding stability properties. The equivalence between geodesic Lyapunov exponents and the damping rates of perturbations bridges linear stability analysis with nonlinear chaos. Building upon these insights, recent studies have proposed using Lyapunov exponents as probes of black hole phase transitions \cite{Guo:2022kio}. At first-order phase transitions, the topological change in spacetime from a small to a large black hole leads to multivalued Lyapunov exponents, each branch corresponding to coexisting phases. Consequently, a sudden jump in the Lyapunov exponent can serve as an order parameter for the transition, displaying mean-field critical exponents. These findings underscore the deep interconnection between the thermodynamic and dynamical aspects of black holes, highlighting how the onset of chaos in particle trajectories or perturbations can reflect the underlying phase structure within the black hole geometry \cite{Yang:2023hci, Lyu:2023sih, Kumara:2024obd, Du:2024uhd, Shukla:2024tkw, Gogoi:2024akv, Chen:2025xqc}.

From both theoretical and holographic perspectives, it is natural to consider black holes carrying non-Abelian charges in addition to (or instead of) the usual Abelian (Maxwell) charge. Non-Abelian gauge fields have self-interactions that significantly differentiate their dynamics from linear Maxwell fields, endowing black holes with distinctive properties. The first analytic black hole solution in Einstein-Yang-Mills (EYM) theory was discovered by Yasskin \cite{Yasskin:1975ag}. Yasskin's solution was subsequently generalized to higher dimensions in Refs. \cite{HabibMazharimousavi:2007fst, Mazharimousavi:2008ap, HabibMazharimousavi:2008zz, HabibMazharimousavi:2008ib} (see Refs. \cite{MasoumiJahromi:2023crl, Naeimipour:2021bgc, Naeimipour:2021dda, Gomez:2023qyv, Rincon:2023hvd, Gomez:2023wei, Gomez:2025pag} for other generalisations and related studies).\footnote{For a comprehensive review of EYM black hole solutions, see Ref. \cite{Hendi:2018sbe}. Studies of EYM black holes in various contexts in asymptotically flat spacetime have notably led to the remarkable discovery of \emph{colored} black holes \cite{Volkov:1989fi, Bizon:1990sr, Kuenzle:1990is, Sudarsky:1992ty, Bjoraker:2000qd, Gao:2003ys}.} When EYM theory is considered within an AdS background, the variety of solutions and their associated stability properties expands further \cite{Radu:2004gu, Winstanley:2008ac, Brihaye:2009cc}. The curvature of AdS spacetime acts as a confining potential, stabilizing field configurations that would otherwise dissipate in asymptotically flat space.

A particularly intriguing extension is the Einstein-Power-Yang-Mills (EPYM) theory \cite{Mazharimousavi:2009mb}. In this model, the action of the non-Abelian gauge field is expressed as a power-law function of the standard Yang-Mills Lagrangian density. This approach is conceptually analogous to employing nonlinear electrodynamics (such as Born-Infeld theory or power-Maxwell invariants) as modifications of Maxwell’s theory - extensions explored to model strong-field corrections and to address singularity structures. Additionally, the EPYM model introduces a tunable parameter that allows exploration of how gauge-field nonlinearity affects black hole solutions. For a generic Yang-Mills charge parameter \( \gamma \), EPYM AdS black holes closely resemble the behavior of Reissner-Nordström AdS black holes, exhibiting van der Waals-type phase transitions when the Yang-Mills charge lies within a specific range \cite{ElMoumni:2018fml, Du:2021xyc, Du:2024amj}. Indeed, numerous studies have demonstrated that the P-V criticality and critical exponents of EPYM AdS black holes frequently coincide with those of conventional charged black holes, and consequently, with the van der Waals fluid \cite{Zhang:2014eap, Yerra:2018mni, Du:2022quq}. This finding suggests a remarkable universality in black hole critical phenomena, extending even to nonlinear non-Abelian matter, which holds intriguing implications for statistical mechanics and dual field theories. Motivated by these intriguing results, in this article we explore the connection between the thermodynamic phase structure of EPYM AdS black holes and the Lyapunov exponents characterizing geodesic stability, thereby correlating thermodynamics and chaotic dynamics associated with gravity coupled to non-Abelian gauge fields.

One of the important methods for studying black hole thermodynamics is the Euclidean path integral formulation of quantum gravity \cite{Gibbons:1977mu, Simovic:2023yuv}. In this framework, one analytically continues the Lorentzian spacetime to imaginary time \( \tau = i t \), enabling the black hole solution to extend smoothly onto a Euclidean manifold with the identification of periodicity in \( \tau \). The Euclidean action \( I_E \) for the gravitational system defines a partition function \( \mathcal{Z} \sim \exp(-I_E) \), analogous to the Boltzmann factor in statistical mechanics. Evaluating \( I_E \) on the black hole solution yields the thermodynamic potential, typically the Gibbs free energy, for the black hole spacetime. Equilibrium thermodynamic quantities, such as entropy and energy, emerge naturally from this partition function through a saddle-point (semiclassical) approximation. For instance, the Hawking temperature arises from the condition that the Euclidean manifold remains regular at the horizon, fixing the periodicity as \( \beta = 1/T \), while the black hole entropy relates directly to the horizon area via the Gibbons-Hawking action term. The Euclidean path integral approach thus provides a first-principles derivation of black hole thermodynamics, with the four thermodynamic laws corresponding to identities satisfied by \( I_E \) and its variations. This approach proves particularly effective in AdS spacetimes, where the ensemble of black hole solutions is well-defined due to the confining boundary conditions of the AdS geometry. In our study, we employ the Euclidean action technique to investigate the thermodynamics of EPYM AdS black holes by explicitly constructing the Euclidean action, including the Yang-Mills field and appropriate boundary terms.

The article is organized as follows: Section \ref{sec:2} discusses the geometry and Euclidean thermodynamics of EPYM AdS black holes. Section \ref{sec:3} analyzes the geodesic motion of both massive and massless particles in unstable orbits, computing the corresponding Lyapunov exponents. In Section \ref{sec:4}, we examine the relationship between Lyapunov exponents and the phase structure of the black hole for timelike and null geodesics. Finally, Section \ref{sec:5} summarizes our results and offers concluding discussions. Additionally, we provide an appendix \ref{app:1} detailing the calculation of the Euclidean action for the EPYM AdS black hole.




\section{Euclidean Thermodynamics of EPYM AdS Black Hole}\label{sec:2}

In this section, we present the thermodynamics of the EPYM black hole in AdS spacetime, utilizing the standard Euclidean approach in canonical ensemble.\footnote{For a brief overview of various methods employed in black hole thermodynamics, see Ref. \cite{Simovic:2023yuv}. For completeness we present the thermodynamics in grand canonical ensemble in \ref{app2}. See Ref. \cite{Braden:1990hw} for a detailed analysis of the Euclidean thermodynamics of charged black holes in grand canonical ensemble.} The action for four- dimensional EPYM gravity is expressed as follows~\cite{Mazharimousavi:2009mb,Zhang:2014eap}: 
\begin{equation}\label{action} 
I = \frac{1}{2} \int d^4 x \sqrt{-g} \left(R - 2\Lambda - \mathcal{F}^\gamma\right), 
\end{equation} 
where $R$ denotes the Ricci scalar, $\Lambda$ is the cosmological constant related to the radius of curvature of the AdS spacetime by $\ell^2=-3/\Lambda$, and $\mathcal{F}$ is defined as: 
\begin{equation} 
\mathcal{F} = \text{Tr}(F_{\mu\nu}^{(a)}F^{(a)\mu\nu}) = \sum_{a=1}^3 F_{\mu\nu}^{(a)}F^{(a)\mu\nu}. 
\end{equation}
The Yang-Mills field strength tensor $F_{\mu\nu}^{(a)}$ is given by:
\begin{equation}
    F_{\mu\nu}^{(a)} = \partial_\mu A_\nu^{(a)} - \partial_\nu A_\mu^{(a)} + \frac{1}{2\xi} C^{(a)}_{(b)(c)} A_\mu^{(b)} A_\nu^{(c)},
\end{equation}
where $C^{(a)}_{(b)(c)}$ are the structure constants of the associated three-parameter Lie group, $\xi$ is the coupling constant, and $A_\mu^{(a)}$ are the gauge potentials corresponding to the $SO(3)$ gauge group \footnote{For detailed discussions on EYM actions associated with the gauge groups SU(2) and SO(3), see Refs. \cite{Oliynyk:2001sk, Volkov:1998cc, Coquereaux:1984ca}.}. Additionally, $\gamma$ is a positive real parameter known as the Yang-Mills charge parameter. In the present analysis, we adopt the Wu-Yang ansatz for the $\mathrm{SO}(3)$ Yang-Mills gauge fields, which ensures a purely magnetic configuration compatible with spherical symmetry \cite{teller1969properties, Yasskin:1975ag, Nakahara:2003nw, Balakin:2015gpq, Balakin:2007nw, Balakin:2006gv}. The ansatz takes the form
\begin{equation}\label{wu_yang}
A^{(a)} = \frac{Q}{r^2} C^{(a)}_{(i)(j)} x^i \, dx^j,
\end{equation}
where $Q$ is the Yang-Mills magnetic charge and $x^i$ are Cartesian spatial coordinates defined by $x^1 = r \sin\theta \cos\varphi$, $x^2 = r \sin\theta \sin\varphi$, and $x^3 = r \cos\theta$. This ansatz leads to a gauge field strength whose invariant takes the form \cite{Stetsko:2020nhi, Stetsko:2020tjg, Chakhchi:2022fls}
\begin{equation}
\mathrm{Tr}(F_{\mu\nu}^{(a)} F^{(a)\mu\nu}) = \frac{Q^2}{r^4}.
\end{equation}

The metric for the EPYM AdS black hole spacetime is given by \cite{Zhang:2014eap}:
\begin{equation}
 \text{d}s^2 = -f(r)\,\text{d}t^2 + \frac{\text{d}r^2}{f(r)} + r^2\left(\text{d}\theta^2 + \sin^2\theta\,\text{d}\varphi^2\right),  
 \label{metric}
\end{equation}
where the metric function $f(r)$ is:
\begin{equation}
 f(r) = 1 - \frac{2M}{r} + \frac{r^2}{\ell^2} + \frac{(2Q^2)^\gamma}{2(4\gamma - 3)r^{4\gamma - 2}}.
 \label{2.2}
\end{equation}
Here, $M$ represents the ADM mass of the black hole, and $Q$ denotes its charge. The weak energy condition restricts the parameter $\gamma$ to satisfy $\gamma \neq \frac{3}{4}$ and $\gamma > 0$ \cite{Mazharimousavi:2009mb}.

Now, we discuss the Euclidean path integral approach to obtain the Euclidean action $I_E$, which is related to the partition function $\mathcal{Z}$. Using this partition function, we can derive the thermodynamic quantities for the EPYM AdS black hole. The Euclidean path integral method connects the partition function $\mathcal{Z}$ of general quantum systems with the Euclidean path integral.

The partition function for a continuous quantum system, characterized by canonical variables $\{q_i\}$ and Hamiltonian $H$ at finite temperature $T = 1/\beta$, is defined as:
\begin{equation}\label{eq:9}
    \mathcal{Z} = \int dq_i \langle q_i | e^{-\beta H} | q_i \rangle.
\end{equation}
Using Eq. (\ref{eq:9}), thermodynamic quantities such as Gibbs free energy ($F$), internal energy ($E$), and entropy ($S$) for the statistical ensemble can be expressed as:
\begin{equation}\label{eq:10}
    F = -T \ln \mathcal{Z}, \quad E = -\frac{\partial \ln \mathcal{Z}}{\partial \beta}, \quad S = -\beta \frac{\partial \ln \mathcal{Z}}{\partial \beta} + \ln \mathcal{Z}.
\end{equation}
In gravitational physics, Gibbons and Hawking pioneered the Euclidean path integral approach \cite{Gibbons:1977mu}, which was later extended by York and collaborators \cite{York:1986it,Braden:1990hw}. Within this context, the partition function can be related to the gravitational path integral \cite{Gibbons:1977mu,York:1986it,Braden:1990hw}:
\begin{equation}\label{eq:11}
    \mathcal{Z} = \int D[g] e^{-I_E[g]} \approx \sum_{g_{\text{cl}}} e^{-I_E[g_{\text{cl}}]},
\end{equation}
where $I_E[g]$ represents the Euclidean action evaluated for the metric $g$, and $I_E[g_{\text{cl}}]$ is the saddle-point (classical) contribution from the metric $g_{\text{cl}}$, which solves the classical equations of motion and satisfies the prescribed boundary conditions.

For a black hole in AdS spacetime, the total Euclidean action in canonical ensemble is expressed as:
\begin{equation}\label{euclidean}
    I_E = I_{\text{bulk}} + I_{\text{surf}} +I_{\text{YM}} + I_{\text{count}},
\end{equation}
where $I_{\text{bulk}}$ is the bulk action derived from Eq. (\ref{action}), $I_{\text{surf}}$ represents the Hawking-Gibbons boundary term, $I_{\text{YM}}$ is an extra boundary term required for canonical ensemble to fix the conserved charge $Q$, and $I_{\text{count}}$ denotes the counter-term included to regularize divergences due to the infinite volume integral of the spacetime \cite{Gibbons:1976ue,Henningson:1998gx,Balasubramanian:1999re,Emparan:1999pm,Caldarelli:1999xj,ELMOUMNI2021115593}. These contributions are explicitly defined as follows:
\begin{equation}\label{eq:13}
    I_{\text{bulk}} = -\frac{1}{16 \pi} \int_{\mathcal{M}} d^4x \sqrt{-g}\left(R - 2\Lambda - \mathcal{F}^\gamma\right),
\end{equation}
\begin{equation}\label{eq:14}
   I_{\text{surf}} = -\frac{1}{8 \pi}\int_{\Sigma_+} d^3x \sqrt{|h|}~ K,
\end{equation}
\begin{equation}
    I_{\text{YM}}=\frac{1}{4\pi}\int_{\Sigma_+}d^3 x\sqrt{h}\ n_\mu \left (\frac{\partial \mathcal{F}^\gamma}{\partial F_{\mu \nu}} \right) A_\nu ,
\end{equation}
\begin{equation}\label{eq:15}
\begin{aligned}
    I_{\text{count}} = & \frac{1}{8 \pi}\int_{\Sigma_+} d^3x \sqrt{|h|}\left[\frac{2}{\ell} + \frac{\ell}{2}\mathcal{R} \right. \\
    & \qquad\qquad \qquad \left. - \frac{\ell^3}{2}\left(\mathcal{R}_{ab}\mathcal{R}^{ab} - \frac{3}{8}\mathcal{R}^2\right)\right],
\end{aligned}
\end{equation}
where $h$ is the determinant of the induced boundary metric $h_{\mu\nu}$, $K$ is the trace of the extrinsic curvature tensor of the boundary hypersurface $\Sigma_+$ embedded within the manifold $\mathcal{M}$, $n$ is an outward point unit normal vector to $\Sigma_+$, and $\mathcal{R}$ and $\mathcal{R}_{ab}$ are, respectively, the Ricci scalar and Ricci tensor computed from the boundary metric $h_{\mu\nu}$.

Integrating equations (\ref{eq:13})-(\ref{eq:15}) and summing them up, we obtain the total Euclidean action as (detailed calculations are provided in \ref{app:1}):
\begin{equation}\label{euclidean_action}
    I_E = \frac{\beta}{2}\left(r_{\text{h}} + \frac{r_{\text{h}}^3}{\ell^2} - \frac{(2Q^2)^\gamma}{2(3 - 4\gamma)r_{\text{h}}^{4\gamma - 3}}\right) - \pi r_{\text{h}}^2.
\end{equation}
Using Eq. (\ref{eq:11}), we have $I_E = -\ln \mathcal{Z}$, from which we can derive the mean thermal energy $E$ as:
\begin{equation}
    E = \frac{\partial I_E}{\partial \beta} = \frac{r_{\text{h}}}{2} + \frac{r_{\text{h}}^3}{2 \ell^2} + \frac{(2Q^2)^\gamma}{4(4\gamma - 3)r_{\text{h}}^{4\gamma - 3}}.
\end{equation}
This expression matches the black hole mass $M$, which is obtained from the horizon condition $f(r_{\text{h}}) = 0$. The equilibrium temperature can be determined by differentiating the action with respect to $r_{\text{h}}$ and solving for $\beta$:
\begin{equation}
    \frac{\partial I_E}{\partial r_{\text{h}}} = 0,
\end{equation}
resulting in the Hawking temperature (which is also the ensemble temperature $T = 1/\beta$):
\begin{equation}\label{temp}
    T = \frac{1}{8\pi}\left(\frac{2}{r_{\text{h}}} + \frac{6 r_{\text{h}}}{\ell^2} - \frac{(2Q^2)^\gamma}{r_{\text{h}}^{4\gamma - 1}}\right).
\end{equation}
The entropy $S$ is obtained as:
\begin{equation}
    S = -\beta \frac{\partial I_E}{\partial \beta} + I_E = \pi r_{\text{h}}^2,
\end{equation}
which exactly matches the Bekenstein-Hawking entropy, $S = A/4$. The generalized free energy is then given by:
\begin{equation}\label{eq:19}
    F = \frac{I_E}{\beta} = \frac{r_{\text{h}}}{4} - \frac{r_{\text{h}}^3}{4\ell^2} + \frac{(2Q^2)^\gamma (4\gamma - 1)}{8(4\gamma - 3)r_{\text{h}}^{4\gamma - 3}},
\end{equation}
which aligns with the Gibbs free energy $F$ computed from $F = M - TS$ \footnote{Note that in the grand-canonical ensemble where the potential $\Phi$ is fixed, the expression for free energy is given by $W = M - T S - \Phi Q$. See \ref{app2}. }. Similarly, the Yang-Mills potential $\Phi$, conjugate to the charge $Q$, is obtained as:
\begin{equation}\label{Phieqn}
    \Phi = \frac{1}{\beta}\frac{\partial I_E}{\partial Q} = \frac{2^{\gamma - 1}Q^{2\gamma - 1}\gamma}{(4\gamma - 3)r_{\text{h}}^{4\gamma - 3}}.
\end{equation}

In the extended phase space, the cosmological constant is treated as a thermodynamic pressure, and the black hole mass is interpreted as the enthalpy. The extended thermodynamics of the EPYM AdS black hole has been previously explored in \cite{Zhang:2014eap,Du:2022quq}. Within this framework, the pressure is defined as:
\begin{equation}
    P = -\frac{\Lambda}{8\pi} = \frac{3}{8\pi \ell^2},
\end{equation}
and the conjugate volume $V$ is given by:
\begin{equation}
    V = -\frac{8\pi}{\beta}\frac{\partial I_E}{\partial \Lambda} = \frac{4\pi r_{\text{h}}^3}{3},
\end{equation}
which corresponds to the geometric volume of the black hole.

Utilizing the above definitions, the thermodynamic quantities in the extended phase space satisfy the first law of thermodynamics:
\begin{equation}
    dM = T dS + \Phi dQ + V dP,
\end{equation}
and the generalized Smarr relation becomes:
\begin{equation}
    M = 2(TS - VP) + \left(\frac{2\gamma - 1}{\gamma}\right) \Phi Q,
\end{equation}
which is obtained through dimensional scaling arguments \cite{Kastor:2009wy,Zhang:2014eap}. For a fixed cosmological constant, the first law simplifies to:
\begin{equation}
    dM = T dS + \Phi dQ.
\end{equation}
Dimensional analysis reveals that physical quantities scale with powers of the AdS radius $\ell$ as follows:
\begin{equation}
\begin{split}
    \tilde{Q} & = \frac{Q}{\ell^{(1 - 2\gamma)/\gamma}}, & \tilde{r}_\text{h} = \frac{r_{\text{h}}}{\ell}, &  \quad \tilde{T} = T\ell, \\
    \quad \tilde{F} & = \frac{F}{\ell}, &  \tilde{M} = \frac{M}{\ell}, & \quad \tilde{r} = \frac{r}{\ell},
    \end{split}
\end{equation}
where the tilde ($\sim$) denotes dimensionless quantities.

Using Eq. (\ref{temp}), the dimensionless horizon radius $\tilde{r}_\text{h}$ can be expressed as a function of the dimensionless temperature $\tilde{T}$. By analyzing the behavior of $\tilde{r}_\text{h}(\tilde{T})$, we can determine the critical charge for a specific value of the parameter $\gamma$. The critical point corresponds to an inflection point, satisfying the following conditions:
\begin{equation}\label{eq:21}
    \frac{\partial \tilde{T}}{\partial \tilde{r}_\text{h}} = 0, \quad \frac{\partial^2 \tilde{T}}{\partial \tilde{r}_\text{h}^2} = 0.
\end{equation}
Solving these conditions simultaneously yields the critical radius, critical charge, and critical temperature. However, analytical solutions for arbitrary values of $\gamma$ are not feasible, necessitating numerical solutions. For example, choosing $\gamma = 3/2$, we obtain the following critical values:
\begin{equation}
    \tilde{r}_{\text{h}c} = 0.4714, \quad \tilde{Q}_c = 0.1325, \quad \tilde{T}_c = 0.2701.
\end{equation}

\begin{figure*}[t]
    \centering
    \includegraphics[width=\textwidth]{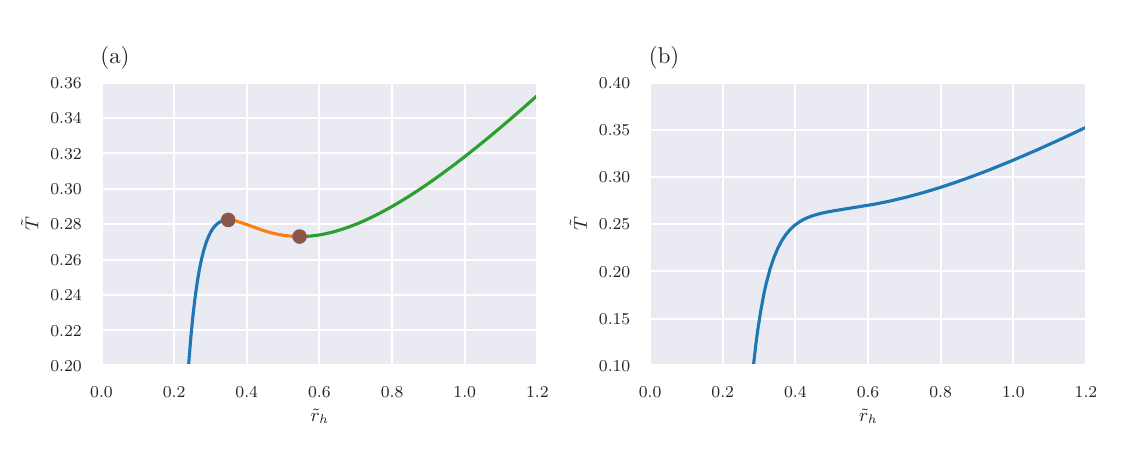}
    \caption{Hawking temperature vs. horizon radius depicted for two distinct charge regimes: $\tilde{Q} = 0.11 < \tilde{Q}_c$ (left) and $\tilde{Q} = 0.16 > \tilde{Q}_c$ (right), and $\gamma =3/2$. 
    }
    \label{fig:1}
\end{figure*}
\begin{figure*}[htp]%
    \centering
     \includegraphics[width=\textwidth]{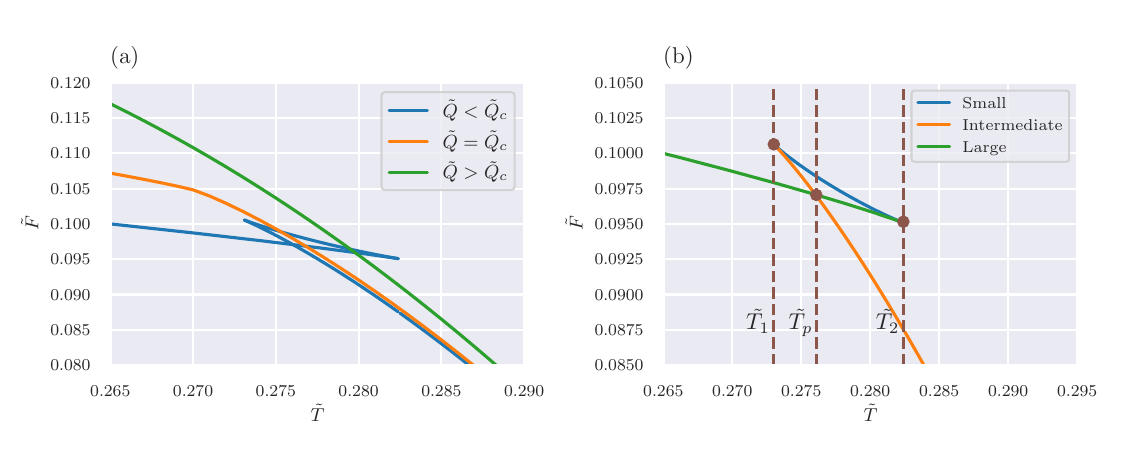}
    \caption{Left: Free energy $\tilde{F}$ as a function of temperature $\tilde{T}$ for three distinct charge regimes: $\tilde{Q}=0.11<\tilde{Q}_c$, $\tilde{Q}= \tilde{Q}_c$ and $\tilde{Q}=0.2>\tilde{Q}_c$, and $\gamma =3/2$. Right: For temperatures in the range $\tilde{T}_1<\tilde{T}<\tilde{T}_2$, three black hole solutions coexist, and a first-order phase transition between small and large black holes occurs at the temperature $\tilde{T}_p$. 
    }%
    \label{fig:2}%
\end{figure*}

As illustrated in Fig.~\ref{fig:1}, for $\tilde{Q}<\tilde{Q}_c$ (left panel), the temperature curve exhibits two turning points (highlighted by brown dots), indicating thermodynamic instability. In contrast, for $\tilde{Q}>\tilde{Q}_c$ (right panel), the absence of turning points signals enhanced stability, with diminishing instability as the charge $\tilde{Q}$ increases. In Fig.~\ref{fig:2}, the free energy $\tilde{F}$ is depicted as a function of the temperature $\tilde{T}$. For $\tilde{Q}<\tilde{Q}_c$, three distinct black hole solutions exist, representing small, intermediate, and large black hole branches. These branches coexist within a specific temperature range, resulting in a first-order phase transition occurring at temperature $\tilde{T}_p$. As the charge $\tilde{Q}$ approaches the critical charge $\tilde{Q}_c$, the intermediate branch gradually disappears, leading the small and large branches to merge, which signals a second-order phase transition at $\tilde{Q}=\tilde{Q}_c$. For charges greater than $\tilde{Q}_c$, the system remains stable, and no further phase transitions occur.




\section{Geodesic Motion and Lyapunov Exponents}\label{sec:3}

In this section, we analyze the relationship between the Lyapunov exponents of massive and massless particles undergoing unstable circular motion near the event horizon of an EPYM AdS black hole and the corresponding phase transitions of these black holes. The Lagrangian describing the geodesics, corresponding to the metric given by Eq. (\ref{metric}), is expressed as:

\begin{equation}\label{eq:24}
    2\mathcal{L} = g_{\mu\nu}\dot{x}^\mu\dot{x}^\nu = -f(r)\dot{t}^2 + \frac{\dot{r}^2}{f(r)} + r^2\dot{\theta}^2 + r^2\sin^2\theta\,\dot{\varphi}^2,
\end{equation}
where dots indicate derivatives with respect to the affine parameter $s$. We focus specifically on unstable circular geodesics confined to the equatorial plane ($\theta = \pi/2$), simplifying the analysis to a two-dimensional phase space. Under this constraint, the Lagrangian reduces to:

\begin{equation}\label{eq:25}
    2\mathcal{L} = -f(r)\dot{t}^2 + \frac{\dot{r}^2}{f(r)} + r^2\dot{\varphi}^2.
\end{equation}

The conjugate momenta $p_\mu = \partial\mathcal{L}/\partial\dot{x}^\mu$ are given by:

\begin{equation}
\begin{aligned}
    p_t = & \frac{\partial\mathcal{L}}{\partial\dot{t}} = -f(r)\dot{t} = -E_n = \text{constant},\\
    p_r = & \frac{\partial\mathcal{L}}{\partial\dot{r}} = \frac{\dot{r}}{f(r)},\\
    p_\varphi = &\frac{\partial\mathcal{L}}{\partial\dot{\varphi}} = r^2\dot{\varphi} = L = \text{constant},
\end{aligned}    
\end{equation}
where $p_r$ is the radial momentum and $L$ represents the angular momentum. The Hamiltonian of the system can then be expressed as:
\begin{equation}\label{hamiltonian}
\begin{aligned}
    2\mathcal{H} &= -f(r)\dot{t}^2 + \frac{\dot{r}^2}{f(r)} + r^2\dot{\varphi}^2 \\[6pt]
    &= \frac{E_n^2}{f(r)} + \frac{\dot{r}^2}{f(r)} + \frac{L^2}{r^2} = \delta_1 = \text{constant},
\end{aligned}
\end{equation}
where $\delta_1 = -1$ for timelike geodesics and $\delta_1 = 0$ for null geodesics. The radial motion of the particle is governed by the equation:
\begin{equation}\label{eq:31}
    \dot{r}^2 + V_{\text{eff}}(r) = 0,
\end{equation}
with the effective potential
\begin{equation}\label{veffeq}
    V_{\text{eff}}(r) = f(r)\left[\frac{L^2}{r^2} + \frac{E_n^2}{f(r)} - \delta_1\right],
\end{equation}
where the constant $E_n$ corresponds to the energy for massless particles or energy per unit mass for massive particles.

Circular orbits with a constant radius are determined by the condition $V'_\text{eff}(r)=0$, and the condition for instability is $V''_\text{eff}(r)<0$. Therefore, the radius of an unstable circular geodesic is found by simultaneously satisfying:
\begin{equation}\label{eq:33}
    V'_\text{eff}(r)=0, \quad V''_\text{eff}(r)<0,
\end{equation}
where primes indicate differentiation with respect to the radial coordinate $r$. Using Eqs. (\ref{hamiltonian}) and (\ref{veffeq}), the Hamiltonian in terms of the effective potential is expressed as:
\begin{equation}
    2\mathcal{H}=\frac{V_\text{eff}(r)}{f(r)}+p_r^2 f(r)+\delta_1.
\end{equation}
This leads to the equations of motion:
\begin{equation}
    \begin{aligned}
        & \dot{r} = \frac{\partial\mathcal{H}}{\partial p_r}=f(r)p_r, \\
        & \dot{p}_r=-\frac{\partial\mathcal{H}}{\partial r}=-\frac{V'_\text{eff}(r)}{2f(r)}+\frac{V_\text{eff}(r) f'(r)}{2f^2(r)}-\frac{p_r^2 f'(r)}{2}.
    \end{aligned}
\end{equation}

Linearizing these equations around the unstable circular orbit at radius $r=r_o$ and applying conditions from Eq. (\ref{eq:33}), we obtain:
\begin{equation}
    \frac{d}{dt}\begin{pmatrix}
    \delta r \\ \delta p_r
    \end{pmatrix}
    = K \begin{pmatrix}
        \delta r \\ \delta p_r
    \end{pmatrix},
\end{equation}
where $K$ is the linear stability matrix \cite{Cornish:2003ig}, given by:
\begin{equation}
  K=  \begin{pmatrix}
        0 & \frac{f(r_o)}{\dot{t}}\\
        -\frac{V_\text{eff}''(r_o)}{2 \dot{t} f(r_o)} & 0
    \end{pmatrix}.
\end{equation}
The principal Lyapunov exponent corresponds to the eigenvalue of the stability matrix $K$ \cite{Cardoso:2008bp}:
\begin{equation}\label{eq:38}
    \lambda = \sqrt{-\frac{V_\text{eff}''(r_o)}{2\dot{t}^2}}.
\end{equation}
Using Eqs. (\ref{eq:31}) and (\ref{veffeq}), the Lyapunov exponent can thus be computed for both massive and massless particle cases.

\subsection{Timelike Geodesics (Massive Particles)}

For EPYM AdS black holes, both stable and unstable circular geodesics exist for massive particles. In this subsection, we specifically focus on unstable timelike circular geodesics characterized by definite angular momentum and energy. For massive particles, $\delta_1=-1$, and using Eq. (\ref{veffeq}), the effective potential is given by:
\begin{equation}\label{eq:39}
    V_\text{eff}(r) = f(r)\left[\frac{L^2}{r^2} + \frac{E_n^2}{f(r)} + 1\right].
\end{equation}
\begin{figure*}[ht]
    \centering
     \includegraphics[width=\textwidth]{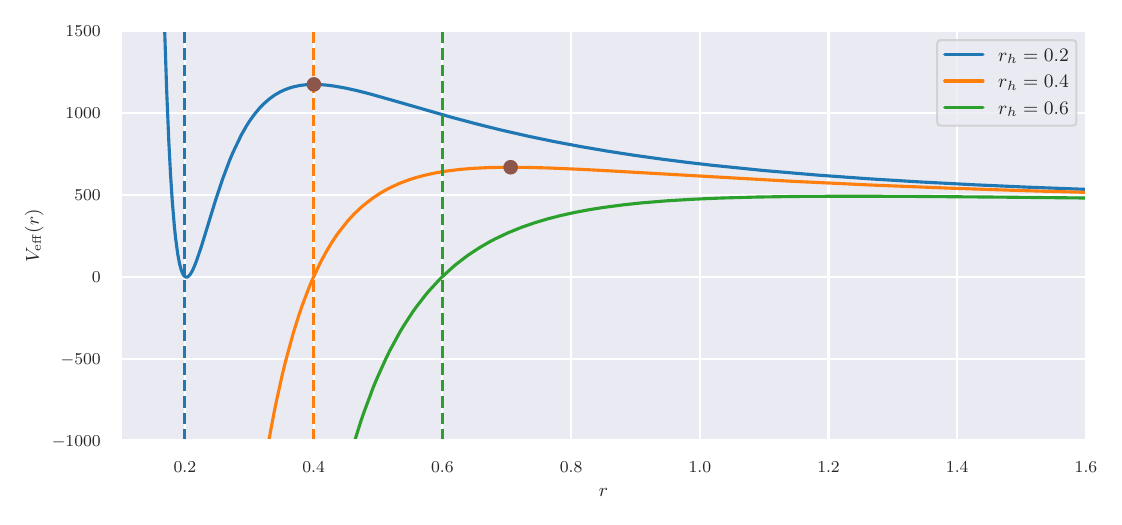}
    \caption{Effective potential $V_\text{eff}$ governing the motion of massive particles around an EPYM AdS black hole, plotted for parameters $\gamma=3/2$, $\tilde{Q}=0.11$, angular momentum $L=20\ell$, and different horizon radii $\tilde{r}_{\text{h}}=0.2$, $0.4$, and $0.6$. The vertical dashed lines represent the event horizon radius, and brown dots indicate unstable circular geodesics. Unstable circular orbits disappear as $\tilde{r}_{\text{h}}$ increases, shown explicitly for $\tilde{r}_{\text{h}}=0.6$.}
    \label{fig:3}
\end{figure*}
Fig.~\ref{fig:3} depicts the effective potential for timelike geodesics across various values of $r_{\text{h}}$, holding other spacetime and particle parameters constant. The plot clearly identifies points of stable and unstable equilibrium. Specifically, the brown dots mark radii corresponding to unstable circular orbits at different $r_{\text{h}}$ values. Notably, these unstable circular orbits disappear with increasing horizon radius $r_{\text{h}}$, signifying a corresponding disappearance of the Lyapunov exponent for larger black holes, as discussed in the next section.

Expressing Eq. (\ref{eq:38}) in terms of the effective potential defined in Eq. (\ref{eq:39}) for timelike geodesics and applying conditions from Eq. (\ref{eq:33}), we have:
\begin{equation}\label{eq:40}
    \lambda = \frac{1}{2}\sqrt{\left(r_o f'(r_o)-2f(r_o)\right)V''_\text{eff}(r_o)}.
\end{equation}
Importantly, this Lyapunov exponent for unstable timelike geodesics remains real and positive, as required by the instability condition $V''_\text{eff}(r_o)<0$.

\subsection{Null Geodesics (Massless Particles)}

For null geodesics, corresponding to massless particles, the effective potential for unstable orbits is obtained by setting $\delta_1 = 0$:
\begin{equation}\label{eq:41}
    V_\text{eff}(r) = f(r)\left[\frac{L^2}{r^2} + \frac{E_n^2}{f(r)}\right].
\end{equation}
Applying the conditions for circular orbit and instability, the Lyapunov exponent for null geodesics is given by:
\begin{equation}\label{eq:42}
    \lambda = \sqrt{-\frac{V''_\text{eff}(r_o)}{2\dot{t}^2}} = \sqrt{-\frac{r_o^2 f'(r_o)}{2L^2}V''_\text{eff}(r_o)}.
\end{equation}
Similar to the massive particle case, the condition $V''_\text{eff}(r_o) < 0$ ensures that $\lambda$ remains positive, indicating instability for null orbits. 



\section{Phase Transition and Lyapunov Exponents}\label{sec:4}

\subsection{Timelike Geodesics (Massive Particles)}

In this subsection, we investigate the phase transition behavior of EPYM AdS black holes by analyzing the Lyapunov exponents of timelike geodesics. From Eq. (\ref{eq:40}), the radius $r_0$ of unstable circular geodesics depends explicitly on the event horizon radius $r_{\text{h}}$, the charge $Q$, and the nonlinear Yang-Mills parameter $\gamma$. Since obtaining an analytical expression for $r_0$ is impractical for arbitrary $\gamma$, we select a fixed value, specifically $\gamma = 3/2$, and numerically solve for $r_0$ using condition \eqref{eq:33}. This approach allows us to examine the variation of the Lyapunov exponent with different values of $r_{\text{h}}$ and $Q$.

\begin{figure*}[t]
    \centering
    \includegraphics[width=\textwidth]{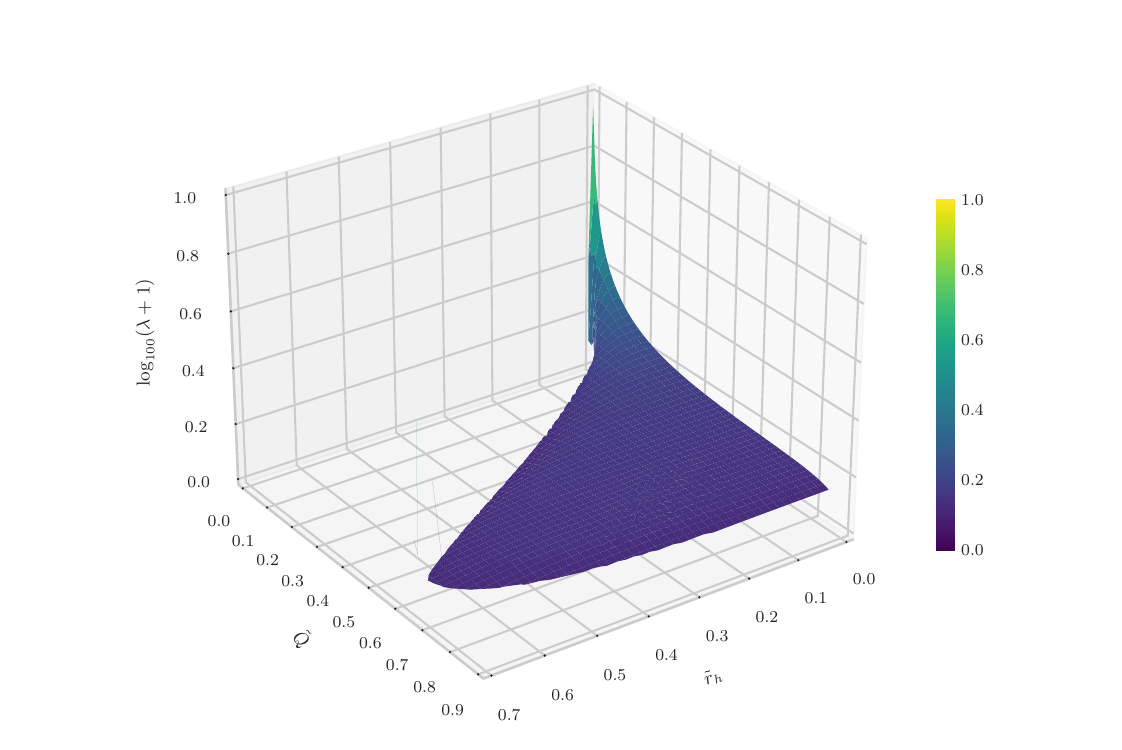}
    \caption{Three-dimensional plot of $\log_{100}(\lambda+1)$ as a function of $\tilde{r}_{\text{h}}$ and $\tilde{Q}$ for a massive particle with angular momentum $L = 20\ell$. Here we chose $\gamma=3/2$. 
    }
    \label{fig:4}
\end{figure*}

The behavior of the Lyapunov exponent $\lambda$ is illustrated in Fig. \ref{fig:4}, where we plot $\log_{100}(\lambda + 1)$ as a function of dimensionless charge $\tilde{Q}$ and dimensionless horizon radius $\tilde{r}_{\text{h}}$ for angular momentum $L = 20\ell$. The figure highlights the divergence of $\lambda$ as $r_{\text{h}} \to 0$. For smaller values of $\tilde{r}_{\text{h}}$, the influence of $\tilde{Q}$ on $\lambda$ is more pronounced compared to larger $\tilde{r}_{\text{h}}$ values. As either $\tilde{r}_{\text{h}}$ or $\tilde{Q}$ increases significantly, their impact on $\lambda$ diminishes, causing $\lambda$ to approach zero.

\begin{figure*}[t]
         \centering
          \includegraphics[width=\textwidth]{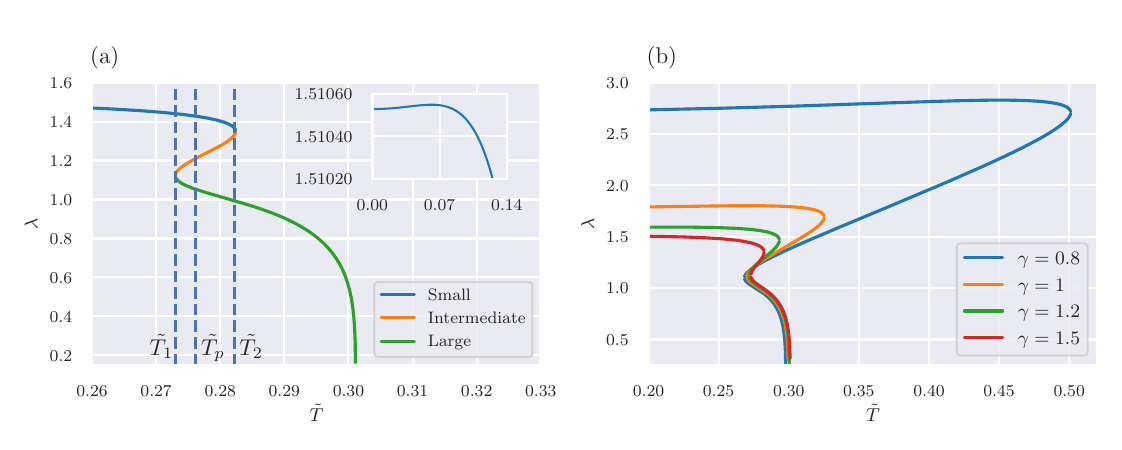}
         \caption{ (a) Lyapunov exponent $\lambda$ for massive particles with angular momentum $L=20\ell$ on unstable timelike circular orbits, plotted as a function of temperature $\tilde{T}$ for $\tilde{Q}=0.11<\tilde{Q}_c$ ($\gamma = 3/2$ case). Three black hole solutions coexist in the temperature range $\tilde{T}_1<\tilde{T}<\tilde{T}_2$, with the phase transition between small and large black holes occurring at $\tilde{T}=\tilde{T}_p$. The inset shows the behavior of $\lambda$ near $\tilde{T}=0$.  
         (b) Lyapunov exponents $\lambda$ of massive particles on unstable timelike circular orbit as a function of temperature $\tilde{T}$ for $\tilde{Q}=0.11<\tilde{Q}_c$ with varying $\gamma$ values. 
         }
         \label{fig:5}  
\end{figure*}

Now, by expressing the Lyapunov exponent as a function of the Hawking temperature, with $\tilde{r}_{\text{h}}$ being a function of $\tilde{T}$, we can illustrate the relationship between $\lambda$ and $\tilde{T}$, as shown in Fig.~\ref{fig:5}. The plot of $\lambda$ versus $\tilde{T}$ closely mirrors the structure seen in the free energy plot (Fig. \ref{fig:2}). Similar to the free energy behavior, the Lyapunov exponent plot exhibits multivalued behavior for $\tilde{Q} < \tilde{Q}_c$ within the temperature range $\tilde{T}_1$ to $\tilde{T}_2$, corresponding to the coexistence of different black hole phases. Specifically, in the small black hole phase, the Lyapunov exponent $\lambda$ initially increases with rising temperature, reaches a maximum, and then decreases as the temperature approaches $\tilde{T}_2$. In contrast, for intermediate and large black hole phases, $\lambda$ respectively increases and decreases with temperature starting from $\tilde{T}_1$. As temperature continues to rise, the Lyapunov exponent asymptotically approaches zero. Conversely, for $\tilde{Q} > \tilde{Q}_c$, no phase transition occurs, consistent with observations from the free energy plots. This analysis clearly demonstrates that Lyapunov exponents as functions of $\tilde{T}$ provide valuable insights into the phase structure of EPYM AdS black holes.

\begin{figure*}[t]
    \centering
     \includegraphics[width=\textwidth]{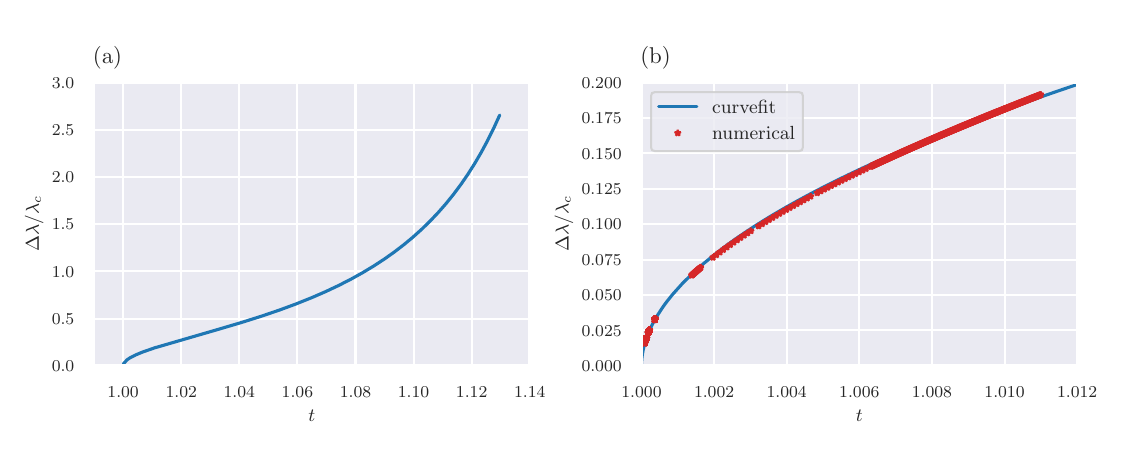}
    \caption{Rescaled discontinuity in Lyapunov exponents plotted as a function of the rescaled phase transition temperature $t = \tilde{T}_p/\tilde{T}_c$ near the critical temperature ($t=1$), for massive particles in unstable timelike circular orbits. Red stars indicate numerical data points, while the solid blue line represents the fitted curve. 
    }
    \label{fig:6}
\end{figure*}

While Fig.~\ref{fig:5} provides qualitative insights into black hole phase transitions, a quantitative analysis is essential to determine order parameters and critical exponents. This quantification is achievable by examining the difference in Lyapunov exponents between small and large black holes. At the first-order phase transition temperature $\tilde{T}_p$, the Lyapunov exponents for the small and large black holes are denoted as $\lambda_s$ and $\lambda_l$, respectively. The difference, defined as $\Delta \lambda = \lambda_s - \lambda_l$, represents a discontinuous jump between phases, serving as an order parameter that remains nonzero during the first-order transition. At the critical point, however, $\lambda_s = \lambda_l$, resulting in $\Delta \lambda = 0$, indicative of a second-order phase transition. Thus, the Lyapunov exponent effectively serves as an order parameter, capturing the critical behavior and characterizing the phase structure of EPYM AdS black holes.

Critical exponents describe the behavior of physical systems near their critical points. To explore the critical behavior of the Lyapunov exponent for timelike geodesics around EPYM AdS black holes, we examine the variation of the normalized difference $\Delta \lambda/\lambda_c$ with respect to the rescaled temperature ratio $\tilde{T}_p/\tilde{T}_c$ (denoted as $t$) near the critical point, as depicted in Fig.~\ref{fig:6}. The left panel illustrates the overall trend, whereas the right panel offers a detailed close-up near the critical region. The critical exponent $b$, associated with the order parameter $\Delta \lambda$, is defined through the relation:
\begin{equation}
    \frac{\Delta \lambda}{\lambda_c} = a (t - 1)^b.
\end{equation}

Numerical analysis reveals the critical exponent to be approximately $b \approx 1/2$. More precisely, the fitted numerical data yield:
\begin{equation}\label{43}
    \frac{\Delta \lambda}{\lambda_c} \approx 4.49463 (t - 1)^{0.527748}.
\end{equation}

This outcome indicates that the critical exponent for $\Delta \lambda$ aligns closely with the critical exponent of the order parameter in the van der Waals fluid model, as anticipated from mean-field theory predictions. Additionally, this result matches the critical exponent determined for null orbits around EPYM AdS black holes \cite{Du:2022quq}.

\subsection{Null Geodesics (Massless Particles)}

In this subsection, we investigate the phase transitions of EPYM AdS black holes using the Lyapunov exponents associated with null geodesics. Our analysis confirms that, regardless of the geodesic type, the relationship between Lyapunov exponents and black hole phase transitions remains consistent, thus demonstrating the robustness of Lyapunov exponents as a probe for phase transitions~\footnote{Since birefringence is expected to occur in non-linear electrodynamics when light propagation is considered, an effective metric is typically introduced to describe such cases~\cite{Hale:2024lzh, Murk:2024nod, Yang:2023hci}. However, as the Power–Yang–Mills field does not couple directly to the electromagnetic sector, birefringence does not arise in our setup in the low-temperature regime that we study. In Yang–Mills theory, birefringence phenomena may instead occur in the propagation of gluon fields~\cite{Lorenci:2008xj}.}.

\begin{figure*}[t]
         \centering
         \includegraphics[width=\textwidth]{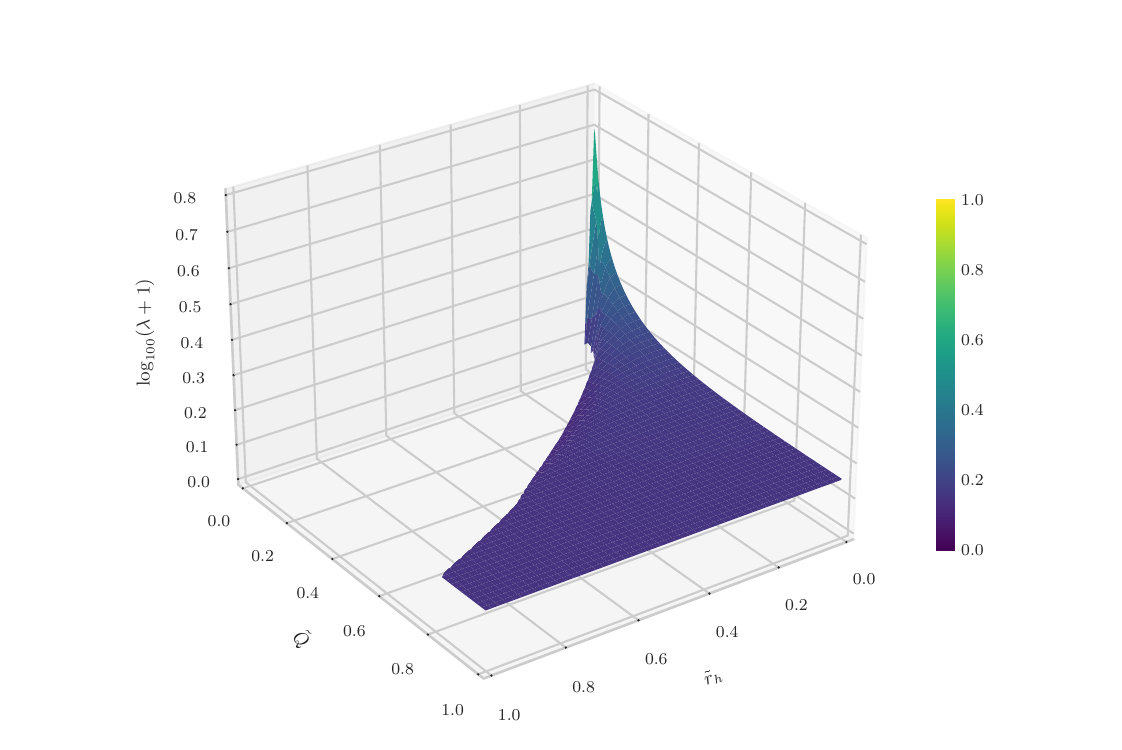}
         \caption{Three-dimensional plot of $\log_{100}(\lambda + 1)$ as a function of $\tilde{r}_{\text{h}}$ and $\tilde{Q}$ for massless particles ($\gamma =3/2$ case). 
         }
         \label{fig:7}
\end{figure*}

Similar to the massive particle scenario, our investigation for massless particles focuses on the specific range of event horizon radii $r_{\text{h}}$ that admit unstable circular geodesics. As in the RN-AdS case \cite{Guo:2022kio}, unstable circular geodesics exist for all angular momenta $L \neq 0$ outside the event horizon. Utilizing Eq. (\ref{eq:42}), we present a three-dimensional plot of $\log_{100}(\lambda + 1)$ as a function of dimensionless charge $\tilde{Q}$ and dimensionless horizon radius $\tilde{r}_{\text{h}}$ in Fig.~\ref{fig:7}. This plot reveals the divergence of $\lambda$ as $\tilde{r}_{\text{h}}$ approaches zero, while $\lambda$ asymptotically approaches one when either $\tilde{Q}$ or $\tilde{r}_{\text{h}}$ becomes very large, consistent with the behavior observed for RN-AdS black holes.

\begin{figure*}[t]
         \centering
         \includegraphics[width=\textwidth]{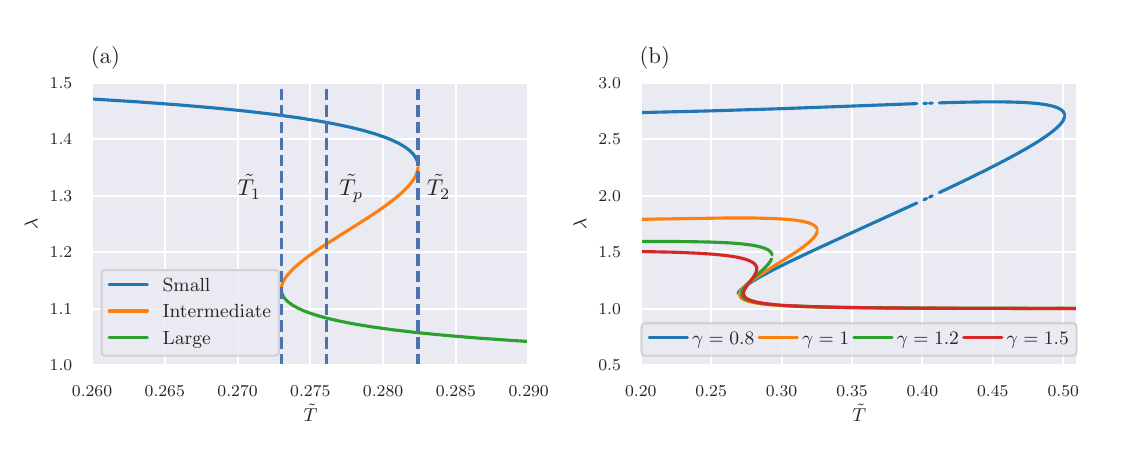}
         \caption{(a). Lyapunov exponent $\lambda$ of massless particles on unstable null circular orbits as a function of temperature $\tilde{T}$ for $\tilde{Q}=0.11<\tilde{Q}_c$. Three black hole phases coexist for temperatures $\tilde{T}_1<\tilde{T}<\tilde{T}_2$, with a phase transition between small and large black holes at $\tilde{T}=\tilde{T}_p$ ($\gamma =3/2$ case). (b). Lyapunov exponents $\lambda$ of massless particles on unstable null circular orbit as a function of temperature $\tilde{T}$ for $\tilde{Q}=0.11<\tilde{Q}_c$ with varying $\gamma$ values. 
         }
         \label{fig:8}
\end{figure*}

Fig.~\ref{fig:8} depicts the Lyapunov exponent $\lambda$ of massless particles on unstable null circular orbits as a function of temperature $\tilde{T}$, obtained by substituting $\tilde{r}_\text{h}(\tilde{T})$ into Eq. (\ref{eq:42}), for a specific charge regime, $\tilde{Q}=0.11<\tilde{Q}_c$. Similar to the massive particle case, unstable null circular geodesics cease to exist beyond a certain temperature, causing $\lambda$ to approach one. Within the temperature range $\tilde{T}_1<\tilde{T}<\tilde{T}_2$, $\lambda$ exhibits multivalued behavior, reflecting the coexistence of three distinct black hole phases, and a first-order phase transition between the small and large black hole phases occurring at temperature $\tilde{T}_p$. Conversely, for $\tilde{Q}>\tilde{Q}_c$, only a single solution exists, with $\lambda$ decreasing monotonically and eventually approaching zero.

\begin{figure*}[t]
    \centering
    \includegraphics[width=\textwidth]{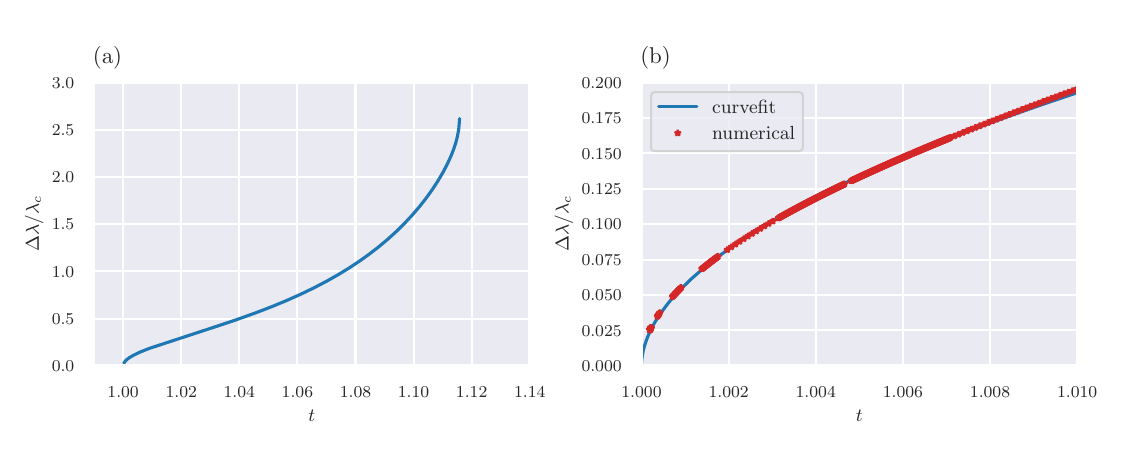}
    \caption{Rescaled discontinuity in Lyapunov exponents plotted as a function of the rescaled phase transition temperature $t = \tilde{T}_p/\tilde{T}_c$, near the critical temperature ($t=1$), for massless particles on unstable null circular geodesics. Red stars indicate numerical data points, while the blue solid line represents the fitted curve. 
    }
    \label{fig:9}%
\end{figure*}

Fig.~\ref{fig:9} further illustrates the discontinuous change in the Lyapunov exponent $\Delta \lambda/\lambda_c$ as a function of the rescaled temperature $\tilde{T}_p/\tilde{T}_c$ (denoted as $t$). This plot clearly indicates that $\Delta \lambda/\lambda_c$ serves effectively as an order parameter for the black hole system. Near the critical temperature, our numerical analysis reveals a critical exponent of approximately $b \approx 1/2$. Specifically, we obtain:
\begin{equation}\label{eq:45}
    \frac{\Delta\lambda}{\lambda_c} \approx 1.97818\,(t - 1)^{0.51967}.
\end{equation}

Thus, we conclude that the critical exponent is independent of both the nonlinear charge of the black hole and the type of geodesic considered, provided the black hole undergoes a phase transition from the small to large black hole phases.




\section{Discussions}\label{sec:5}

In this article, we have investigated the Euclidean thermodynamics of Einstein-Power-Yang-Mills AdS (EPYM AdS) black holes and used the Lyapunov exponent to probe their phase structure. The relationship between black hole phase transitions and chaos, characterized by the Lyapunov exponents, is explored by considering both timelike and null geodesics in the vicinity of EPYM AdS black holes. Our study generalizes the connection between Lyapunov exponents and black hole phase structures from Abelian gauge field (Maxwell) coupled gravity spacetimes to spacetimes coupled with nonlinear and non-Abelian gauge fields.

First, we presented the thermodynamic properties of the EPYM AdS black hole by calculating the on-shell Euclidean action. We generalized the approach typically used for conventional gauge fields coupled to Einstein gravity to incorporate Power-Yang-Mills gauge fields coupled to Einstein gravity, including the appropriate boundary and counter terms. The thermodynamic variables of the black hole obtained from the Euclidean action satisfy the first law of thermodynamics and the Smarr relation. The black hole phase transition was studied by examining the resulting free energy as a function of Hawking temperature. A transition between small and large black hole phases, analogous to the conventional van der Waals-like system, was observed. Due to the nonlinear Yang-Mills charge parameter ($\gamma$), analytical analysis was not feasible for arbitrary values of $\gamma$; hence, our analysis was carried out numerically.

The underlying phase structure of the black hole spacetime is examined by calculating the Lyapunov exponents for both timelike and null geodesics, offering a complementary perspective. Our results effectively demonstrate that Lyapunov exponents serve as a robust framework for characterizing black hole phase structures. Specifically, we showed that Lyapunov exponents display distinct thermal behaviors that mirror the relevant free energies when system parameters remain below their critical values. Within a certain temperature range, Lyapunov exponents exhibit multivalued behavior, reflecting the coexistence of three distinct black hole solutions. Above the critical value, however, the Lyapunov exponent becomes monotonic across the entire temperature range. The transition from multivalued to single-valued behavior of the Lyapunov exponent accurately signifies the second-order critical point, highlighting that the thermodynamic properties and phase structures of EPYM AdS black holes are encoded within these exponents.

Further, we demonstrated that the discontinuity in Lyapunov exponents ($\Delta \lambda$) during the first-order phase transition serves as an effective order parameter, capturing the essence of the black hole phase transition. Notably, we identified that $\Delta \lambda$ possesses a critical exponent of $1/2$ at the critical point, further reinforcing the analogy with van der Waals phase transitions. Although variations in the $\gamma$ parameter influence critical parameter values, the overall behavior of Lyapunov exponents and associated critical exponents remains consistent.

Recently, the Euclidean path integral approach to black hole thermodynamics and phase transitions has attracted renewed interest \cite{Liu:2023sbf, Li:2024hje}. As future work, it would be valuable to explore the kinetics of black hole phase transitions and generalized free energy landscapes using the Euclidean path integral framework for EPYM AdS black holes. Additionally, since the Euclidean approach has been applied to nonlinear Maxwell Lagrangian densities coupled to Einstein gravity in (Anti) de-Sitter backgrounds \cite{Simovic:2023yuv, Soranidis:2023cyd}, it would be intriguing to extend this method to EPYM black hole spacetimes.

\section{Acknowledgment}
ANK's research was supported by the Croatian Science Foundation Project No. IP-2020-02-9614 \textit{Search for Quantum spacetime in Black Hole QNM spectrum and Gamma Ray Bursts}.

\appendix

\section{Euclidean Action Calculation}\label{app:1}

In this appendix, we provide detailed calculations of the Euclidean action \eqref{euclidean_action}. Recall that the action for EPYM gravity in four-dimensional spacetime, including a cosmological constant $\Lambda$, is given by Eq. \eqref{action}:
\begin{equation}\label{5.1}
   I = \frac{1}{2}\int_{\mathcal{M}} d^4x \sqrt{-g}\left(R - 2\Lambda - \mathcal{F}^\gamma\right),
\end{equation}
with the metric:
\begin{equation}\label{5.2}
    \text{d}s^2 = -f(r)\,\text{d}t^2 + \frac{\text{d}r^2}{f(r)} + r^2\text{d}\theta^2 + r^2\sin^2\theta\,\text{d}\varphi^2,
\end{equation}
\noindent where
\begin{equation}\label{5.3}
    f(r) = 1 + \frac{r^2}{\ell^2} - \frac{2M}{r} + \frac{(2Q^2)^\gamma}{2(4\gamma - 3)r^{4\gamma - 2}}.
\end{equation}

The Euclidean action is expressed by Eq. \eqref{euclidean} as:
\begin{equation}\label{5.4}
    I_E = I_{\text{bulk}} + I_{\text{surf}}+I_{\text{YM}} + I_{\text{count}}.
\end{equation}
\noindent We calculate the Euclidean action for an arbitrary temperature $T$, equivalently described by the imaginary time period $\beta$. The bulk action contribution to the total Euclidean action is:
\begin{equation}\label{5.8}
    \begin{aligned}
         I_{\text{bulk}} &= -\frac{1}{16 \pi} \int_{\mathcal{M}} d^4x \sqrt{-g}\left(R - 2\Lambda - \mathcal{F}^\gamma\right)\\
         &= -\frac{1}{16\pi} \int_0^\beta d\tau \int_{r_{\text{h}} +\epsilon}^{\tilde{R}} dr \int_0^\pi d\theta \\
         & \qquad \times \int_0^{2\pi} d\varphi \left[r^2 \sin\theta \left(R - \frac{6}{\ell^2} - \frac{(2Q^2)^\gamma}{r^{4\gamma}}\right)\right] \\
         &= \frac{\beta}{2}\left(\frac{(\tilde{R}^3 + r_{\text{h}}^3)}{\ell^2} - M - r_{\text{h}}\frac{(2Q^2)^\gamma}{4(3 - 4\gamma)r_{\text{h}}^{4\gamma - 3}}\right) - \pi r_{\text{h}}^2,
    \end{aligned}
\end{equation}
where we assume that the boundary $\Sigma_+$ is positioned at $r = \tilde{R}$, introducing a cutoff in the integral at this boundary. Here, we have identified the periodicity in $\tau$ with the Killing surface gravity $\beta^{-1}=f'(r_h)/4\pi~$\cite{Simovic:2023yuv}. Eventually, we take the limit $\tilde{R}\to \infty$.

To calculate the surface term $I_{\text{surf}}$, we first identify the non-zero components of the induced boundary metric $h_{\mu\nu}$:
\begin{equation}\label{5.9}
    \begin{aligned}
        h_{\tau\tau} &= f(r)|_{r=\tilde{R}},\\
        h_{\theta\theta}&= \tilde{R}^2,\\
        h_{\varphi\varphi} &= \tilde{R}^2 \sin^2{\theta}.
    \end{aligned}
\end{equation}
To find the trace of extrinsic curvature $K$, we use the outward unit normal vector $n^\mu = (0, \sqrt{f(r)}, 0, 0)$, leading to:
\begin{equation}\label{5.10}
   \begin{aligned}
        K & = h^{\mu\nu}K_{\mu\nu} = h^{\mu\nu}\nabla_\mu n_\nu = h^{\mu\nu}(\partial_\mu n_\nu - \Gamma^\rho_{\mu\nu} n_\rho)\\
        & = -\left(h^{\tau\tau}\Gamma^r_{\tau\tau} + h^{\theta\theta}\Gamma^r_{\theta\theta} + h^{\varphi\varphi}\Gamma^r_{\varphi\varphi}\right)n_r\Big|_{r=\tilde{R}}\\
        & = \frac{1}{\sqrt{f(\tilde{R})}} \left[ \frac{2}{\tilde{R}} - \frac{3M}{\tilde{R}^2} + \frac{3 \tilde{R}}{\ell^2} - \frac{(2Q^2)^\gamma}{2(4\gamma - 3)} (3 - 2\gamma) \tilde{R}^{1 - 4\gamma} \right].
   \end{aligned}
\end{equation}
Using Eqs. (\ref{5.9}) and (\ref{5.10}), we calculate the surface term as:
\begin{equation}\label{5.11}
\begin{aligned}
    I_{\text{surf}} &= -\frac{1}{8 \pi }\int _{\Sigma_+} d^3x\sqrt{|h|}\, K\\
    &= -\frac{\beta}{2} \left( 2 \tilde{R} + \frac{3 \tilde{R}^3}{\ell^2} - 3M + \frac{(2Q^2)^\gamma}{4\gamma - 3} (3 - 2\gamma) \tilde{R}^{3 - 4\gamma} \right).
\end{aligned}
\end{equation}
In the canonical ensemble to fix the conserved charge $Q$, we must add an extra boundary term to the action, which is defined as~\cite{ELMOUMNI2021115593},
\begin{equation}
    I_{\text{YM}}=\frac{1}{4\pi}\int_{\Sigma_+}d^3 x\sqrt{h}\ n_\mu \left (\frac{\partial \mathcal{F}^\gamma}{\partial F_{\mu \nu}} \right) A_\nu ,
\end{equation}
For the Wu-Yang magnetic ansatz \ref{wu_yang}, $A_0=0$, and there is no electric potential at the boundary, since the only non-vanishing components of $F_{\mu\nu}$ are $F_{23}=-F_{32}$. Hence, this term will vanish.

Next, the counter-term $I_{\text{count}}$ can be evaluated as:
\begin{widetext}
\begin{equation}\label{5.12}
\begin{aligned}
        I_{\text{count}} &= \frac{1}{8 \pi} \int_{\Sigma_+} d^3x\sqrt{|h|}\left[\frac{2}{\ell} + \frac{\ell}{2} \mathcal{R} - \frac{\ell^3}{2} \left( \mathcal{R}_{ab} \mathcal{R}^{ab} - \frac{3}{8} \mathcal{R}^2 \right)\right]\\
        &= \frac{1}{8\pi} \int_0^\beta d\tau \int_0^\pi d\theta \int_0^{2\pi} d\varphi \sqrt{1 + \frac{\tilde{R}^2}{\ell^2} -\frac{2 M}{\tilde{R}} +\frac{(2Q^2)^\gamma}{2 (4\gamma -3)\tilde{R}^{4\gamma -2}}}\,\tilde{R}^2 \sin\theta \left(\frac{2}{\ell}+\frac{\ell}{\tilde{R}^2}-\frac{\ell^2}{4\tilde{R}^4}\right) \\
        &= \beta \left( \frac{\tilde{R}^3}{\ell^2} + \tilde{R} - M \right),
\end{aligned}
\end{equation}
where we have employed a Taylor expansion:
\begin{equation}\label{5.13}
    \sqrt{1 + \frac{\tilde{R}^2}{\ell^2} -\frac{2 M}{\tilde{R}} +\frac{(2Q^2)^\gamma}{2 (4\gamma -3)\tilde{R}^{4\gamma -2}}}=\frac{\tilde{R}}{\ell}\left[1-\frac{\ell^2M}{\tilde{R}^3}+\frac{\ell^2}{2\tilde{R}^2}-\frac{\ell^4}{8\tilde{R}^4}+\mathcal{O}\left(\frac{1}{\tilde{R}^4}\right)\right],
\end{equation}
\end{widetext}
and omitted terms vanishing as $\tilde{R}\to\infty$. Assuming $\gamma > 3/4$ (as the surface term diverges at $\gamma = 3/4$), the total Euclidean action for the Einstein-Power-Yang-Mills AdS black hole becomes:
\begin{equation}\label{5.14}
    I_E= \frac{\beta}{2}\left(r_{\text{h}}+\frac{r_{\text{h}}^3}{\ell^2}-\frac{(2Q^2)^\gamma}{2(3-4\gamma)r_{\text{h}}^{4\gamma-3}}\right)-\pi r_{\text{h}}^2.
\end{equation}

\section{Thermodynamics in grand canonical ensemble}\label{app2}
While the thermodynamics and phase transitions of EPYM black holes have been extensively analyzed in the canonical ensemble, studies in the grand canonical ensemble remain comparatively limited. Investigating these black holes in the grand canonical framework is important not only for clarifying their internal thermodynamic structure, but also for applications within the AdS/CFT correspondence, where black hole phase transitions are often linked to those of the dual quark–gluon plasma under a fixed boundary potential. Fixing the potential at the AdS boundary provides a more natural description of the dual plasma than the fixed–charge (canonical) ensemble. Here we examine the phase transitions of EPYM black holes in the grand canonical ensemble, employing both free energy and Lyapunov exponents as probes of critical behavior in this exotic class of solutions.

The grand canonical free energy $W$ can be defined by a Legendre transform of the canonical free energy $F$,
\begin{equation}
    W = F - \Phi Q = M - TS - \Phi Q.
\end{equation}
which gives
\begin{equation}\label{eq:WG}
     W=\frac{r_{\text{h}}}{4}-\frac{r_{\text{h}}^3}{4\ell^2}+\frac{(2Q^2)^\gamma(4-5\gamma)}{8(4\gamma-3)r_{\text{h}}^{4\gamma-3}}
\end{equation}
In the grand canonical ensemble, the Yang-Mills potential $\Phi$ is fixed rather than the charge $Q$.
Hence, we can rewrite the equation \eqref{Phieqn} for $Q$ in terms of $\Phi$. We get,
\begin{equation}
    Q=\left(\frac{1}{\gamma}\frac{(4\gamma-3)\Phi}{2^{\gamma-1} r^{4\gamma-3}}\right)^{\frac{1}{2\gamma-1}}
\end{equation}
Substituting this in (\ref{eq:WG}), we can rewrite the free energy $W$ in terms of $\Phi$.
\begin{figure*}[htp]
    \centering
    \includegraphics[width=\textwidth]{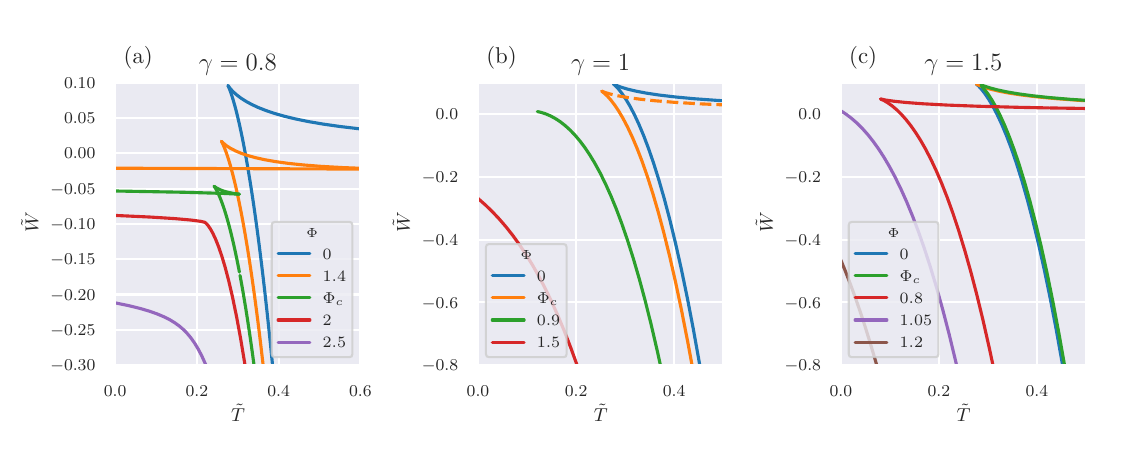}
    \caption{Free Energy $\tilde{W}$ vs Temperature $\tilde{T}$ for varying $\gamma$ values in the grand canonical ensemble for different $\Phi$ values. The $\Phi_{c}$ values in (a) is 1.746, (b) is 0.4028 and (c) is 0.118.
    }
    \label{fig:ap1}
\end{figure*}
Now, from Fig. \ref{fig:ap1}, we study the phase structure and stability of the EPYM black hole by observing the variation of free energy as a function of the temperature. The cusp in the $W$–$T$ diagram signals a first-order phase transition. Interestingly, the characteristic swallow-tail structure, indicative of a van der Waals–type phase transition, emerges only for specific values of the parameter $\gamma$, as shown here for $\gamma=0.8$. In contrast, for $\gamma=1$ and $\gamma=1.5$, such behavior disappears. Previous studies in the grand canonical ensemble have demonstrated that van der Waals–type phase transitions generally occur only for black holes in $d \geq 5$ dimensions \cite{Dehghani:2019thq}, with exceptions arising in certain non-trivial cases such as Power-Maxwell and Born–Infeld black holes \cite{Liang:2019dni,Hendi:2012um,Hale:2024lzh}. Our results reveal that, for particular values of $\gamma$, black holes in the Power–Yang–Mills class can also exhibit van der Waals–type phase transitions in four dimensions within the grand canonical ensemble, thereby adding them to this family of non-trivial solutions. For $\gamma \geq 1$, the van der Waals behavior ceases to exist, and the system instead undergoes a Hawking–Page–type phase transition.

\begin{figure*}[htp]
    \centering
    \includegraphics[width=\textwidth]{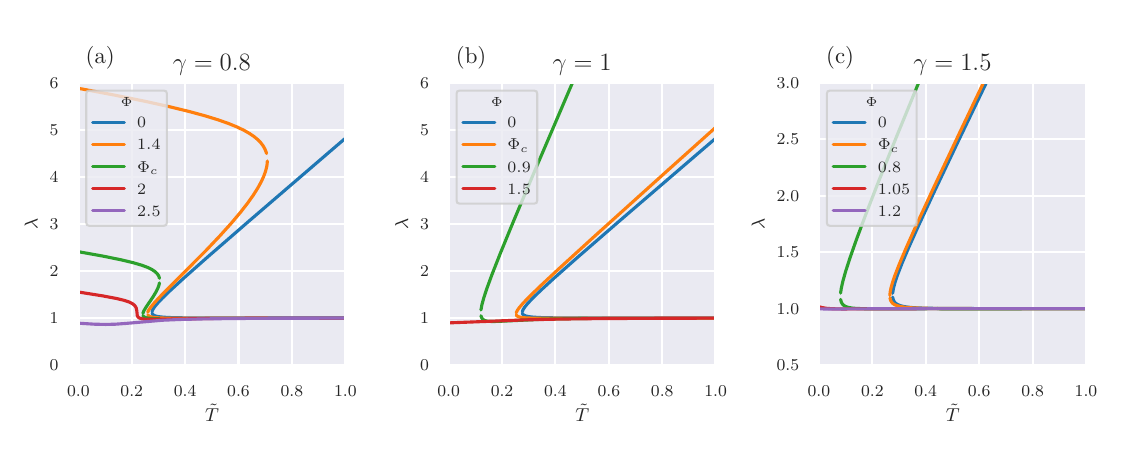}
    \caption{Lyapunov exponent $\lambda$ vs Temperature $\tilde{T}$ for varying $\gamma$ values in the grand canonical ensemble for different $\Phi$ values. The $\Phi_{c}$ values are same as in Fig. \ref{fig:ap1}.
    }
    \label{fig:ap2}
\end{figure*}

Similar to the canonical ensemble, the Lyapunov exponents of massless particles orbiting the black hole can be employed to probe phase transitions. The variation of the Lyapunov exponent with temperature as shown in Fig. \ref{fig:ap2}, reflects the behavior observed in  Fig \ref{fig:ap1}. 


\bibliographystyle{spphys}       
\bibliography{BibTex}

\begin{thebibliography}{10}
\providecommand{\url}[1]{{#1}}
\providecommand{\urlprefix}{URL }
\expandafter\ifx\csname urlstyle\endcsname\relax
  \providecommand{\doi}[1]{DOI \discretionary{}{}{}#1}\else
  \providecommand{\doi}{DOI \discretionary{}{}{}\begingroup \urlstyle{rm}\Url}\fi

\bibitem{Witten:2024upt}
E.~Witten, Eur. Phys. J. Plus \textbf{140}(5), 430 (2025).
\newblock \doi{10.1140/epjp/s13360-025-06288-y}

\bibitem{Hawking:1971tu}
S.W. Hawking, Phys. Rev. Lett. \textbf{26}, 1344 (1971).
\newblock \doi{10.1103/PhysRevLett.26.1344}

\bibitem{Bekenstein:1972tm}
J.D. Bekenstein, Lett. Nuovo Cim. \textbf{4}, 737 (1972).
\newblock \doi{10.1007/BF02757029}

\bibitem{Bekenstein:1973ur}
J.D. Bekenstein, Phys. Rev. D \textbf{7}, 2333 (1973).
\newblock \doi{10.1103/PhysRevD.7.2333}

\bibitem{Bardeen:1973gs}
J.M. Bardeen, B.~Carter, S.W. Hawking, Commun. Math. Phys. \textbf{31}, 161 (1973).
\newblock \doi{10.1007/BF01645742}

\bibitem{Hawking:1974rv}
S.W. Hawking, Nature \textbf{248}, 30 (1974).
\newblock \doi{10.1038/248030a0}

\bibitem{Hawking:1975vcx}
S.W. Hawking, Commun. Math. Phys. \textbf{43}, 199 (1975).
\newblock \doi{10.1007/BF02345020}.
\newblock [Erratum: Commun.Math.Phys. 46, 206 (1976)]

\bibitem{Hawking:1982dh}
S.W. Hawking, D.N. Page, Commun. Math. Phys. \textbf{87}, 577 (1983).
\newblock \doi{10.1007/BF01208266}

\bibitem{Witten:1998zw}
E.~Witten, Adv. Theor. Math. Phys. \textbf{2}, 505 (1998).
\newblock \doi{10.4310/ATMP.1998.v2.n3.a3}

\bibitem{Chamblin:1999hg}
A.~Chamblin, R.~Emparan, C.V. Johnson, R.C. Myers, Phys. Rev. D \textbf{60}, 104026 (1999).
\newblock \doi{10.1103/PhysRevD.60.104026}

\bibitem{Chamblin:1999tk}
A.~Chamblin, R.~Emparan, C.V. Johnson, R.C. Myers, Phys. Rev. D \textbf{60}, 064018 (1999).
\newblock \doi{10.1103/PhysRevD.60.064018}

\bibitem{Caldarelli:1999xj}
M.M. Caldarelli, G.~Cognola, D.~Klemm, Class. Quant. Grav. \textbf{17}, 399 (2000).
\newblock \doi{10.1088/0264-9381/17/2/310}

\bibitem{Kastor:2009wy}
D.~Kastor, S.~Ray, J.~Traschen, Class. Quant. Grav. \textbf{26}, 195011 (2009).
\newblock \doi{10.1088/0264-9381/26/19/195011}

\bibitem{Dolan:2011xt}
B.P. Dolan, Class. Quant. Grav. \textbf{28}, 235017 (2011).
\newblock \doi{10.1088/0264-9381/28/23/235017}

\bibitem{Kubiznak:2012wp}
D.~Kubiznak, R.B. Mann, JHEP \textbf{07}, 033 (2012).
\newblock \doi{10.1007/JHEP07(2012)033}

\bibitem{Gunasekaran:2012dq}
S.~Gunasekaran, R.B. Mann, D.~Kubiznak, JHEP \textbf{11}, 110 (2012).
\newblock \doi{10.1007/JHEP11(2012)110}

\bibitem{Kubiznak:2016qmn}
D.~Kubiznak, R.B. Mann, M.~Teo, Class. Quant. Grav. \textbf{34}(6), 063001 (2017).
\newblock \doi{10.1088/1361-6382/aa5c69}

\bibitem{Altamirano:2013ane}
N.~Altamirano, D.~Kubiznak, R.B. Mann, Phys. Rev. D \textbf{88}(10), 101502 (2013).
\newblock \doi{10.1103/PhysRevD.88.101502}

\bibitem{Altamirano:2013uqa}
N.~Altamirano, D.~Kubiz\v{n}\'ak, R.B. Mann, Z.~Sherkatghanad, Class. Quant. Grav. \textbf{31}, 042001 (2014).
\newblock \doi{10.1088/0264-9381/31/4/042001}

\bibitem{Frassino:2014pha}
A.M. Frassino, D.~Kubiznak, R.B. Mann, F.~Simovic, JHEP \textbf{09}, 080 (2014).
\newblock \doi{10.1007/JHEP09(2014)080}

\bibitem{NaveenaKumara:2019nnt}
A.~Naveena~Kumara, C.L. Ahmed~Rizwan, S.~Punacha, K.M. Ajith, M.S. Ali, Phys. Rev. D \textbf{102}(8), 084059 (2020).
\newblock \doi{10.1103/PhysRevD.102.084059}

\bibitem{NaveenaKumara:2020biu}
A.~Naveena~Kumara, C.L. Ahmed~Rizwan, K.~Hegde, M.S. Ali, K.M. Ajith, Phys. Rev. D \textbf{103}(4), 044025 (2021).
\newblock \doi{10.1103/PhysRevD.103.044025}

\bibitem{doi:10.1080/00207179208934253}
A.M. LYAPUNOV, International Journal of Control \textbf{55}(3), 531 (1992).
\newblock \doi{10.1080/00207179208934253}

\bibitem{Maldacena:2015waa}
J.~Maldacena, S.H. Shenker, D.~Stanford, JHEP \textbf{08}, 106 (2016).
\newblock \doi{10.1007/JHEP08(2016)106}

\bibitem{Zhao:2018wkl}
Q.Q. Zhao, Y.Z. Li, H.~Lu, Phys. Rev. D \textbf{98}(12), 124001 (2018).
\newblock \doi{10.1103/PhysRevD.98.124001}

\bibitem{Guo:2020pgq}
X.~Guo, K.~Liang, B.~Mu, P.~Wang, M.~Yang, Chin. Phys. C \textbf{45}(2), 023115 (2021).
\newblock \doi{10.1088/1674-1137/abcf20}

\bibitem{Gwak:2022xje}
B.~Gwak, N.~Kan, B.H. Lee, H.~Lee, JHEP \textbf{09}, 026 (2022).
\newblock \doi{10.1007/JHEP09(2022)026}

\bibitem{Cardoso:2008bp}
V.~Cardoso, A.S. Miranda, E.~Berti, H.~Witek, V.T. Zanchin, Phys. Rev. D \textbf{79}(6), 064016 (2009).
\newblock \doi{10.1103/PhysRevD.79.064016}

\bibitem{Guo:2021enm}
G.~Guo, P.~Wang, H.~Wu, H.~Yang, JHEP \textbf{06}, 060 (2022).
\newblock \doi{10.1007/JHEP06(2022)060}

\bibitem{Guo:2022kio}
X.~Guo, Y.~Lu, B.~Mu, P.~Wang, JHEP \textbf{08}, 153 (2022).
\newblock \doi{10.1007/JHEP08(2022)153}

\bibitem{Yang:2023hci}
S.~Yang, J.~Tao, B.~Mu, A.~He, JCAP \textbf{07}, 045 (2023).
\newblock \doi{10.1088/1475-7516/2023/07/045}

\bibitem{Lyu:2023sih}
X.~Lyu, J.~Tao, P.~Wang, Eur. Phys. J. C \textbf{84}(9), 974 (2024).
\newblock \doi{10.1140/epjc/s10052-024-13354-9}

\bibitem{Kumara:2024obd}
A.N. Kumara, S.~Punacha, M.S. Ali, JCAP \textbf{07}, 061 (2024).
\newblock \doi{10.1088/1475-7516/2024/07/061}

\bibitem{Du:2024uhd}
Y.Z. Du, H.F. Li, Y.B. Ma, Q.~Gu, Eur. Phys. J. C \textbf{85}(1), 78 (2025).
\newblock \doi{10.1140/epjc/s10052-025-13809-7}

\bibitem{Shukla:2024tkw}
B.~Shukla, P.P. Das, D.~Dudal, S.~Mahapatra, Phys. Rev. D \textbf{110}(2), 024068 (2024).
\newblock \doi{10.1103/PhysRevD.110.024068}

\bibitem{Gogoi:2024akv}
N.J. Gogoi, S.~Acharjee, P.~Phukon,   (2024)

\bibitem{Chen:2025xqc}
D.~Chen, C.~Yang, Y.~Liu, Phys. Lett. B \textbf{865}, 139463 (2025).
\newblock \doi{10.1016/j.physletb.2025.139463}

\bibitem{Yasskin:1975ag}
P.B. Yasskin, Phys. Rev. D \textbf{12}, 2212 (1975).
\newblock \doi{10.1103/PhysRevD.12.2212}

\bibitem{HabibMazharimousavi:2007fst}
S.~Habib~Mazharimousavi, M.~Halilsoy, Phys. Rev. D \textbf{76}, 087501 (2007).
\newblock \doi{10.1103/PhysRevD.76.087501}

\bibitem{Mazharimousavi:2008ap}
S.H. Mazharimousavi, M.~Halilsoy, Phys. Lett. B \textbf{659}, 471 (2008).
\newblock \doi{10.1016/j.physletb.2007.11.006}

\bibitem{HabibMazharimousavi:2008zz}
S.~Habib~Mazharimousavi, M.~Halilsoy, JCAP \textbf{12}, 005 (2008).
\newblock \doi{10.1088/1475-7516/2008/12/005}

\bibitem{HabibMazharimousavi:2008ib}
S.~Habib~Mazharimousavi, M.~Halilsoy, Phys. Lett. B \textbf{665}, 125 (2008).
\newblock \doi{10.1016/j.physletb.2008.06.007}

\bibitem{MasoumiJahromi:2023crl}
F.~Masoumi~Jahromi, B.~Mirza, F.~Naeimipour, S.~Nasirimoghadam, Nucl. Phys. B \textbf{993}, 116271 (2023).
\newblock \doi{10.1016/j.nuclphysb.2023.116271}

\bibitem{Naeimipour:2021bgc}
F.~Naeimipour, B.~Mirza, F.~Masoumi~Jahromi, Eur. Phys. J. C \textbf{81}(5), 455 (2021).
\newblock \doi{10.1140/epjc/s10052-021-09241-2}

\bibitem{Naeimipour:2021dda}
F.~Naeimipour, B.~Mirza, S.~Nasirimoghadam, Phys. Rev. D \textbf{104}(10), 104059 (2021).
\newblock \doi{10.1103/PhysRevD.104.104059}

\bibitem{Gomez:2023qyv}
G.~G\'omez, A.~Rinc\'on, N.~Cruz, Annals Phys. \textbf{459}, 169489 (2023).
\newblock \doi{10.1016/j.aop.2023.169489}

\bibitem{Rincon:2023hvd}
A.~Rincon, G.~G\'omez, Phys. Dark Univ. \textbf{46}, 101576 (2024).
\newblock \doi{10.1016/j.dark.2024.101576}

\bibitem{Gomez:2023wei}
G.~G\'omez, J.F. Rodr\'\i{}guez, Phys. Rev. D \textbf{108}(2), 024069 (2023).
\newblock \doi{10.1103/PhysRevD.108.024069}

\bibitem{Gomez:2025pag}
G.~Gomez, J.F. Rodriguez, Eur. Phys. J. C \textbf{85}(8), 921 (2025).
\newblock \doi{10.1140/epjc/s10052-025-14657-1}

\bibitem{Hendi:2018sbe}
S.H. Hendi, M.~Momennia, Phys. Lett. B \textbf{777}, 222 (2018).
\newblock \doi{10.1016/j.physletb.2017.12.033}

\bibitem{Volkov:1989fi}
M.S. Volkov, D.V. Galtsov, JETP Lett. \textbf{50}, 346 (1989)

\bibitem{Bizon:1990sr}
P.~Bizon, Phys. Rev. Lett. \textbf{64}, 2844 (1990).
\newblock \doi{10.1103/PhysRevLett.64.2844}

\bibitem{Kuenzle:1990is}
H.P. Kuenzle, A.K.M. Masood-ul Alam, J. Math. Phys. \textbf{31}, 928 (1990).
\newblock \doi{10.1063/1.528773}

\bibitem{Sudarsky:1992ty}
D.~Sudarsky, R.M. Wald, Phys. Rev. D \textbf{46}, 1453 (1992).
\newblock \doi{10.1103/PhysRevD.46.1453}

\bibitem{Bjoraker:2000qd}
J.~Bjoraker, Y.~Hosotani, Phys. Rev. D \textbf{62}, 043513 (2000).
\newblock \doi{10.1103/PhysRevD.62.043513}

\bibitem{Gao:2003ys}
S.~Gao, Phys. Rev. D \textbf{68}, 044016 (2003).
\newblock \doi{10.1103/PhysRevD.68.044016}

\bibitem{Radu:2004gu}
E.~Radu, E.~Winstanley, Phys. Rev. D \textbf{70}, 084023 (2004).
\newblock \doi{10.1103/PhysRevD.70.084023}

\bibitem{Winstanley:2008ac}
E.~Winstanley, Lect. Notes Phys. \textbf{769}, 49 (2009).
\newblock \doi{10.1007/978-3-540-88460-6_2}

\bibitem{Brihaye:2009cc}
Y.~Brihaye, E.~Radu, D.H. Tchrakian, Phys. Rev. D \textbf{81}, 064005 (2010).
\newblock \doi{10.1103/PhysRevD.81.064005}

\bibitem{Mazharimousavi:2009mb}
S.H. Mazharimousavi, M.~Halilsoy, Phys. Lett. B \textbf{681}, 190 (2009).
\newblock \doi{10.1016/j.physletb.2009.10.006}

\bibitem{ElMoumni:2018fml}
H.~El~Moumni, Phys. Lett. B \textbf{776}, 124 (2018).
\newblock \doi{10.1016/j.physletb.2017.11.037}

\bibitem{Du:2021xyc}
Y.Z. Du, H.F. Li, F.~Liu, R.~Zhao, L.C. Zhang, Chin. Phys. C \textbf{45}(11), 112001 (2021).
\newblock \doi{10.1088/1674-1137/ac2049}

\bibitem{Du:2024amj}
Y.Z. Du, H.H. Zhao, Y.~Zhang, Q.~Gu,   (2024)

\bibitem{Zhang:2014eap}
M.~Zhang, Z.Y. Yang, D.C. Zou, W.~Xu, R.H. Yue, Gen. Rel. Grav. \textbf{47}(2), 14 (2015).
\newblock \doi{10.1007/s10714-015-1851-2}

\bibitem{Yerra:2018mni}
P.K. Yerra, B.~Chandrasekhar, Mod. Phys. Lett. A \textbf{34}(27), 1950216 (2019).
\newblock \doi{10.1142/S021773231950216X}

\bibitem{Du:2022quq}
Y.Z. Du, H.F. Li, F.~Liu, L.C. Zhang, JHEP \textbf{01}, 137 (2023).
\newblock \doi{10.1007/JHEP01(2023)137}

\bibitem{Gibbons:1977mu}
G.W. Gibbons, S.W. Hawking, Phys. Rev. D \textbf{15}, 2738 (1977).
\newblock \doi{10.1103/PhysRevD.15.2738}

\bibitem{Simovic:2023yuv}
F.~Simovic, I.~Soranidis, Phys. Rev. D \textbf{109}(4), 044029 (2024).
\newblock \doi{10.1103/PhysRevD.109.044029}

\bibitem{Braden:1990hw}
H.W. Braden, J.D. Brown, B.F. Whiting, J.W. York, Jr., Phys. Rev. D \textbf{42}, 3376 (1990).
\newblock \doi{10.1103/PhysRevD.42.3376}

\bibitem{Oliynyk:2001sk}
T.A. Oliynyk, H.P. Kunzle, Class. Quant. Grav. \textbf{19}, 457 (2002).
\newblock \doi{10.1088/0264-9381/19/3/303}

\bibitem{Volkov:1998cc}
M.S. Volkov, D.V. Gal'tsov, Phys. Rept. \textbf{319}, 1 (1999).
\newblock \doi{10.1016/S0370-1573(99)00010-1}

\bibitem{Coquereaux:1984ca}
R.~Coquereaux, A.~Jadczyk, Commun. Math. Phys. \textbf{98}, 79 (1985).
\newblock \doi{10.1007/BF01211045}

\bibitem{teller1969properties}
E.~Teller, H.~Mark, S.~Fernbach, \emph{Properties of Matter Under Unusual Conditions: In Honor of Edward Teller's 60th Birthday} (Interscience Publishers, 1969)

\bibitem{Nakahara:2003nw}
M.~Nakahara, \emph{{Geometry, topology and physics}} (2003)

\bibitem{Balakin:2015gpq}
A.B. Balakin, J.P.S. Lemos, A.E. Zayats, Phys. Rev. D \textbf{93}(2), 024008 (2016).
\newblock \doi{10.1103/PhysRevD.93.024008}

\bibitem{Balakin:2007nw}
A.B. Balakin, H.~Dehnen, A.E. Zayats, Phys. Rev. D \textbf{76}, 124011 (2007).
\newblock \doi{10.1103/PhysRevD.76.124011}

\bibitem{Balakin:2006gv}
A.B. Balakin, A.E. Zayats, Phys. Lett. B \textbf{644}, 294 (2007).
\newblock \doi{10.1016/j.physletb.2006.12.002}

\bibitem{Stetsko:2020nhi}
M.M. Stetsko, Phys. Rev. D \textbf{101}(12), 124017 (2020).
\newblock \doi{10.1103/PhysRevD.101.124017}

\bibitem{Stetsko:2020tjg}
M.M. Stetsko, Gen. Rel. Grav. \textbf{53}(1), 2 (2021).
\newblock \doi{10.1007/s10714-020-02777-w}

\bibitem{Chakhchi:2022fls}
L.~Chakhchi, H.~El~Moumni, K.~Masmar, Phys. Rev. D \textbf{105}(6), 064031 (2022).
\newblock \doi{10.1103/PhysRevD.105.064031}

\bibitem{York:1986it}
J.W. York, Jr., Phys. Rev. D \textbf{33}, 2092 (1986).
\newblock \doi{10.1103/PhysRevD.33.2092}

\bibitem{Gibbons:1976ue}
G.W. Gibbons, S.W. Hawking, Phys. Rev. D \textbf{15}, 2752 (1977).
\newblock \doi{10.1103/PhysRevD.15.2752}

\bibitem{Henningson:1998gx}
M.~Henningson, K.~Skenderis, JHEP \textbf{07}, 023 (1998).
\newblock \doi{10.1088/1126-6708/1998/07/023}

\bibitem{Balasubramanian:1999re}
V.~Balasubramanian, P.~Kraus, Commun. Math. Phys. \textbf{208}, 413 (1999).
\newblock \doi{10.1007/s002200050764}

\bibitem{Emparan:1999pm}
R.~Emparan, C.V. Johnson, R.C. Myers, Phys. Rev. D \textbf{60}, 104001 (1999).
\newblock \doi{10.1103/PhysRevD.60.104001}

\bibitem{ELMOUMNI2021115593}
H.~{El Moumni}, J.~Khalloufi, Nuclear Physics B \textbf{973}, 115593 (2021).
\newblock \doi{https://doi.org/10.1016/j.nuclphysb.2021.115593}

\bibitem{Cornish:2003ig}
N.J. Cornish, J.J. Levin, Class. Quant. Grav. \textbf{20}, 1649 (2003).
\newblock \doi{10.1088/0264-9381/20/9/304}

\bibitem{Hale:2024lzh}
T.~Hale, D.~Kubiznak, J.~Men{\v{s}}{\'\i}kov{\'a}, Phys. Rev. D \textbf{109}(8), 084061 (2024).
\newblock \doi{10.1103/PhysRevD.109.084061}

\bibitem{Murk:2024nod}
S.~Murk, I.~Soranidis, Phys. Rev. D \textbf{110}(4), 044064 (2024).
\newblock \doi{10.1103/PhysRevD.110.044064}

\bibitem{Lorenci:2008xj}
V.A.D. Lorenci, S.Y. Li, Phys. Rev. D \textbf{78}, 034004 (2008).
\newblock \doi{10.1103/PhysRevD.78.034004}

\bibitem{Liu:2023sbf}
C.~Liu, R.~Li, K.~Zhang, J.~Wang, JHEP \textbf{11}, 068 (2023).
\newblock \doi{10.1007/JHEP11(2023)068}

\bibitem{Li:2024hje}
R.~Li, C.~Liu, J.~Wang, Phys. Rev. D \textbf{110}(2), 024079 (2024).
\newblock \doi{10.1103/PhysRevD.110.024079}

\bibitem{Soranidis:2023cyd}
I.~Soranidis, Phys. Rev. D \textbf{109}(4), 044041 (2024).
\newblock \doi{10.1103/PhysRevD.109.044041}

\bibitem{Dehghani:2019thq}
A.~Dehghani, S.H. Hendi, Class. Quant. Grav. \textbf{37}(2), 024001 (2020).
\newblock \doi{10.1088/1361-6382/ab5eb4}

\bibitem{Liang:2019dni}
K.~Liang, P.~Wang, H.~Wu, M.~Yang, Eur. Phys. J. C \textbf{80}(3), 187 (2020).
\newblock \doi{10.1140/epjc/s10052-020-7750-z}

\bibitem{Hendi:2012um}
S.H. Hendi, M.H. Vahidinia, Phys. Rev. D \textbf{88}(8), 084045 (2013).
\newblock \doi{10.1103/PhysRevD.88.084045}

\end{thebibliography}
\nocite{}
\end{document}